\documentclass[10pt,journal,compsoc]{IEEEtran}
\usepackage{times}  %Required
\usepackage{helvet}  %Required
\usepackage{color}
\usepackage{courier}  %Required
\usepackage{url}  %Required
\usepackage{graphicx}  %Required
\usepackage{amsmath}
\usepackage{amsthm}
\usepackage{amsfonts}
\usepackage{subfigure}
\usepackage{multirow}
\usepackage{algorithm}
\usepackage{algorithmic}
\usepackage[normalem]{ulem}
\useunder{\uline}{\ul}{}
\usepackage{booktabs}

\graphicspath{ {imgs/} }

\theoremstyle{definition}
\newtheorem{defn}{Definition}

\def\b1{\mathbf{1}}
\def\mG{\mathcal{G}}
\def\mA{\mathcal{A}}
\def\mE{\mathcal{E}}
\def\mU{\mathcal{U}}
\def\mM{\mathcal{M}}
\def\be{\mathbf{e}}
\def\bx{\mathbf{x}}
\def\by{\mathbf{y}}
\def\bv{\mathbf{v}}

\def\bB{\mathbf{B}}
\def\bP{\mathbf{P}}

\def\bC{\mathbf{C}}
\def\bH{\mathbf{H}}
\def\bU{\mathbf{U}}
\def\bX{\mathbf{X}}
\def\bY{\mathbf{Y}}
\def\bZ{\mathbf{Z}}
\def\bW{\mathbf{W}}

\def\et{~et.al. }

\begin{document}
\title{Ranking Users in Social Networks with Motif-based PageRank}
\author{Huan Zhao \IEEEmembership{Member,~IEEE} and Xiaogang Xu and Yangqiu Song \IEEEmembership{Member,~IEEE} and Dik Lun Lee and Zhao Chen and Han Gao\\
\IEEEcompsocitemizethanks{\IEEEcompsocthanksitem Huan Zhao, Yangqiu Song and Dik Lun Lee are with the Department of Computer Science and Engineering,
	Hong Kong University of Science and Technology, Clear Water Bay,
	Hong Kong.
	E-mail: \{hzhaoaf, yqsong, dlee\}@cse.ust.hk.

\IEEEcompsocthanksitem Xiaogang Xu is with College of Information Science \&  Electronic Engineering, Zhejiang University, Hangzhou, China. E-mail: \mbox{xiaogangxu@zju.edu.cn.}

\IEEEcompsocthanksitem Zhao Chen and Han Gao are with Tencent Technology (SZ) Co., Ltd., China.
E-mail: \{gilbertchen,alangao\}@tencent.com.

\IEEEcompsocthanksitem Huan Zhao and Xiaogang Xu contribute equally to this work.
}
}

\IEEEtitleabstractindextext{
\begin{abstract}
PageRank has been widely used to measure the authority or the influence of a user in social networks. However, conventional PageRank only makes use of edge-based relations, which represent first-order relations between two connected nodes. It ignores higher-order relations that may exist between nodes.
In this paper, we propose a novel framework, motif-based PageRank (MPR), to incorporate higher-order relations into the conventional PageRank computation. Motifs are subgraphs consisting of a small number of nodes. We use motifs to capture higher-order relations between nodes in a network and introduce two methods, one linear and one non-linear, to combine PageRank with higher-order relations. We conduct extensive experiments on three real-world networks, namely, DBLP, Epinions, and Ciao. We study different types of motifs, including 3-node simple and anchor motifs, 4-node and 5-node motifs. Besides using single motif, we also run MPR with ensemble of multiple motifs. We also design a learning task to evaluate the abilities of authority prediction with motif-based features. All experimental results demonstrate that MPR can significantly improve the performance of user ranking in social networks compared to the baseline methods.
\end{abstract}

}
\maketitle

\begin{figure*}
	\centering
	\subfigure[Example graph.]{\includegraphics[width=0.2\textwidth]{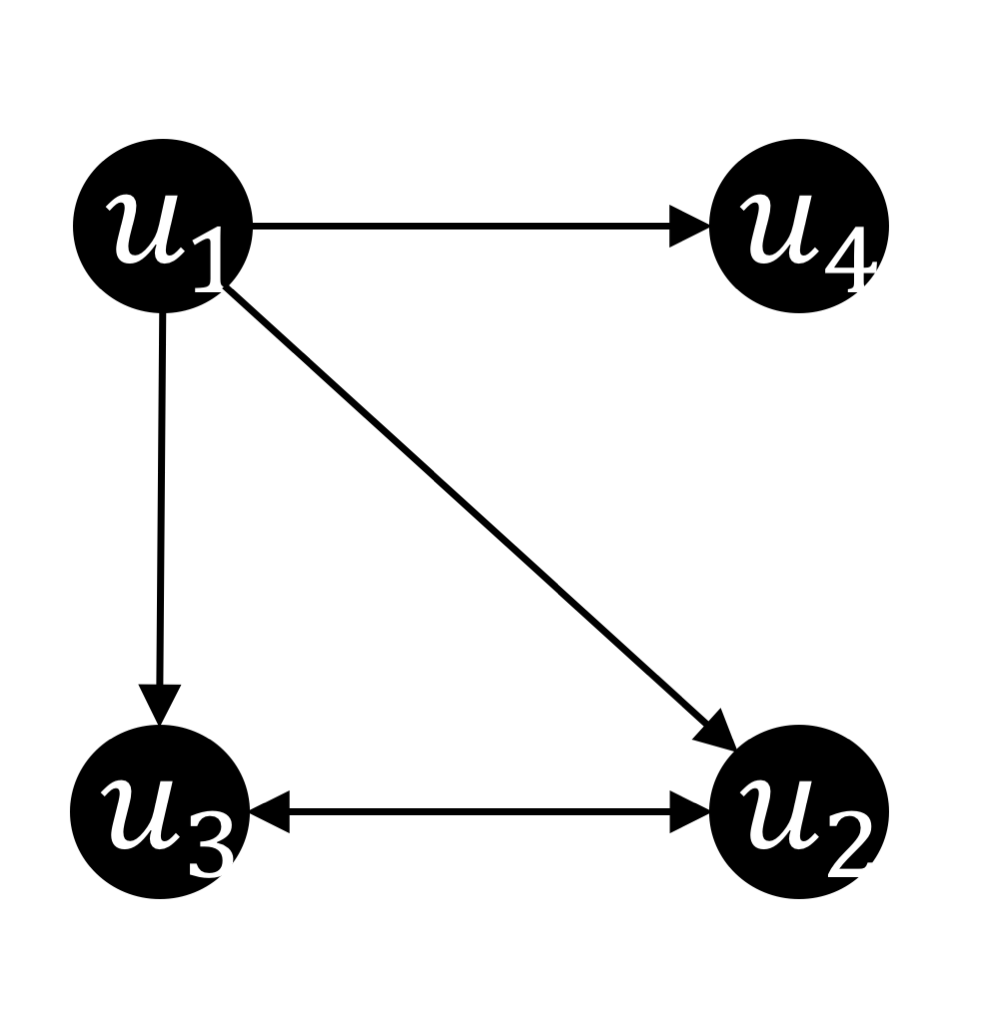}}\vspace{0.1in}
	\subfigure[Triangular motifs.]{\includegraphics[width=0.7\textwidth]{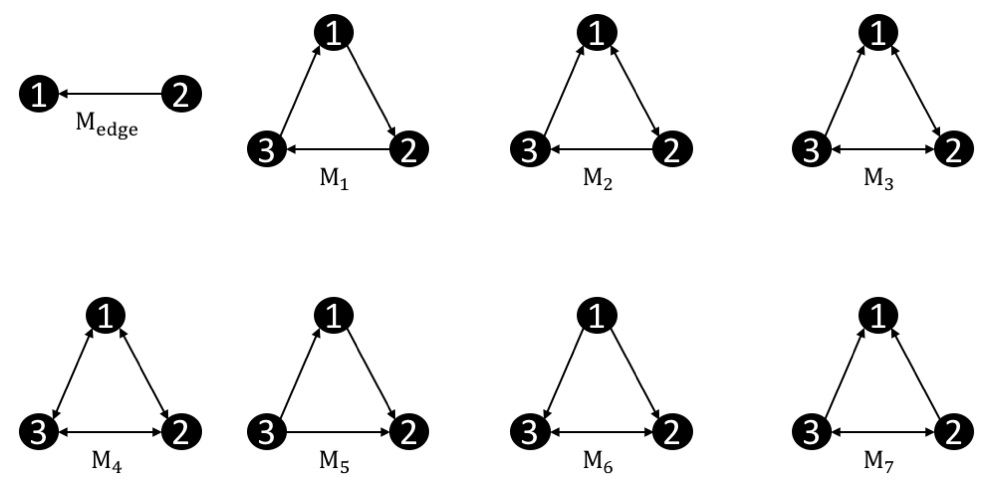}}
	\caption{(a) is an directed graph with 4 nodes. (b) includes all triangular motifs in a directed unweighted graph. Note that we use $M_{edge}$ to represent the edge-based relation, and indices $1,2,3$ are used to represent the position of a node in a motif. Note that there is no strict rules to name a motif, and we adopt the names for triangular motifs from~\cite{benson2016higher}.}
	\label{fig-motif-example}  
\end{figure*}

\section{Introduction}
Online social networks have become a ubiquitous platform for people to share their opinions and thus exert influence on each other. Through analyzing a social network, we can rank users for different purposes, e.g.,
finding opinion leaders~\cite{SongCHT07}, influential users~\cite{TangSWY09,xiang2013pagerank}, and trustworthiness of users~\cite{WangWTZC15}. 
PageRank~\cite{page1999pagerank} has been very successful in computing the authority of webpages in a webgraph, and have been demonstrated to be useful for ranking users in social networks~\cite{xing2004weighted,radicchi2009diffusion,zyczkowski2010citation,ding2011applying,xiang2013pagerank}. The original PageRank~\cite{page1999pagerank} is applied to unweighed graph, which assumes all outgoing links from a node have the same weight, and weighted PageRank, i.e., assigning different weights to different outgoing links from a node,  have been shown more effective for ranking researchers in citation networks~\cite{xing2004weighted,radicchi2009diffusion,zyczkowski2010citation,ding2011applying}.

Despite the success of the PageRank algorithm in ranking users in social network, there is a major problem facing it, including the binary, i.e,, original, and weighted PageRank algorithms. That is, the weights are calculated based on the direct relations between two nodes, e.g., the frequency of a researcher citing another one in a citation network, or assigning larger weights to a link pointing by a more influential user in Twitter. In this paper, the direct relation between two nodes are termed as \textit{first-order} relation. Then, the problem is that existing PageRank algorithms only consider the first-order relations between two nodes, while ignoring \textit{higher-order} relations captured by local structures, or called \textit{network motif}, which have been demonstrated important in complex networks~\cite{milo2002network,benson2016higher}.

%However, there is a major problem facing the original PageRank algorithm. That is, it assumes that all outgoing links from a node have the same weight. Thus, a node is influenced equally by all of the nodes it directly connects to. However, this is not the case in real life, as, for example, in Twitter, a user follows different users for different reasons and thus should receive different influences from the followees. In this paper, direct links are referred to as \textit{first-order} relations, and PageRank is simple and effective in authority computation based on first-order relations. In this paper, we argue that while first-order relations are essential in authority computation, we should also consider \textit{higher-order} relations captured by local structures, or called \textit{network motif}~\cite{milo2002network,benson2016higher}.

We use a concrete example to illustrate this problem. Figure~\ref{fig-motif-example}(a) depicts a social trust network, where a link from user $u_i$ to user $u_j$ means that $u_i$ trusts $u_j$. Considering direct links alone, PageRank assumes that $u_2$, $u_3$ and $u_4$ have the same influence on $u_1$ because the direct links are unweighted.
However, when considering the local structure, i.e., the triangle containing $u_1$, $u_2$, and $u_3$, we can observe that comparing to $u_4$, $u_2$ and $u_3$ should have larger influence on $u_1$. This is because $u_2$ and $u_3$ have mutual influence on each other, and in addition to the direct links, their influence on $u_1$ can also pass through the indirect links. This phenomenon is  consistent with the observation that trianglular relations are indicator of strong social relations~\cite{simmel1908sociology,granovetter1977strength}.

In this paper, we argue that the weights of the links between directly connected users should be adjusted based on the local structures they participate in. To achieve this, we propose a novel framework, called Motif-based PageRank (MPR), for the task of user ranking, which combines both first-order and higher-order relations in a social network. In Figure~\ref{fig-motif-example}(b), we show seven typical 3-node motifs, in which $M_6$ characterizes the triangular structure in the above example. Note that we also refer to the first-order relations as edge-based relations, denoted by $M_{edge}$ in Figure~\ref{fig-motif-example}(b). In MPR, we first compute the motif-based adjacency matrix, which captures the pairwise relations between two nodes in a specific motif. Then, we combine the motif-based adjacency matrix with the edge-based adjacency matrix to re-weigh the links between users. Thus, we can incorporate higher-order relations into authority computation performed by the conventional PageRank algorithm. 

Preliminary results of this paper have been reported in~\cite{zhao2018ranking}. 
In this full version, in additional to 3-node simple motifs, we also study the performance of other types of motifs, including 3-node anchor motifs, 4-node and 5-node simple motifs, which are discussed in Section~\ref{sec-framework}.
We propose different algorithms to compute the motif-based adjacency matrices. Besides, we propose a non-linear combination method to integrate edge-based and motif-based relations. Furthermore, we design an ensemble method to incorporate multiple motifs simultaneously. Finally, we conduct extensive experiments to demonstrate the effectiveness of the above methods.
Experiments on an academic citation dataset, DBLP, and two trustworthiness datasets, Epinions and Ciao, to extract influential or trustworthy users by the number of users who trust them.
The results show that motif-based and edge-based relations are complementary to each other in computing the authority of a user in a social network.
MPR significantly outperforms conventional PageRank and other baselines for the task of user ranking in social networks. Moreover, we design a learning task of authority prediction in social networks, which also demonstrate the usefulness of motif-based features. The code of this work is available at \url{https://github.com/HKUST-KnowComp/Motif-based-PageRank}.

The rest of the paper is organized as follows. In Section~\ref{sec-rel}, we first review the existing works on PageRank and motif analysis in complex networks. Then we introduce in detail MPR in Section~\ref{sec-framework}. In Section~\ref{sec-exp}, we further present our experimental results as well as the analysis. In Section~\ref{sec-learning-exp}, we design a learning task to evaluate the abilities of authority prediction with motif-based features. Finally, we conclude our work in Section~\ref{sec-conclusion}.

\section{Related Work}
\label{sec-rel}
In this section, we introduce related work on authority computation with PageRank and motif in complex networks.

\subsection{PageRank}
PageRank was first introduced to rank webpages on the Internet~\cite{page1999pagerank}.
Apart from ranking webpages, PageRank has been used in many other domains~\cite{gleich2015pagerank}, such as citation network analysis~\cite{ding2011topic} and link prediction~\cite{liben2007link}. 
In~\cite{xiang2013pagerank}, Xiang\et explicitly connected PageRank with social influence model and showed that authority is equivalent to influence under their framework. 
Thus, PageRank can also help to select influential nodes in networks. 
Moreover, PageRank has been used to identify opinion leaders~\cite{SongCHT07} and find trustworthy users~\cite{WangWTZC15} in social networks. Besides, weighted PageRank has been demonstrated effective for ranking researchers in citation networks~\cite{xing2004weighted,radicchi2009diffusion,zyczkowski2010citation,ding2011applying}.
Compared to our work, all of the previous studies only considered direct edges in PageRank computation and ignored higher-order relations among multiple nodes.

\subsection{Motif in Complex Networks}
Motif characterizes higher-order relations in complex networks and is also associated with other names such as graphlets or subgraphs.
Network motif was first introduced in~\cite{milo2002network}.
It has been shown to be useful in many applications such as social networks~\cite{ugander2013subgraph,granovetter1973strength,rotabi2017detecting}, scholar networks~\cite{wang2014identification}, biology~\cite{prvzulj2007biological}, neuroscience~\cite{sporns2004motifs}, and temporal networks~\cite{paranjape2017motifs}. 
Besides, most of the previous work focused on how to efficiently count the number of motifs in complex networks~\cite{ahmed2015efficient,jha2015path,wang2016minfer,han2016waddling,stefani2017triest,pinar2017escape}. 
Recently, it was proven that motifs can also be used for graph clustering or community detection~\cite{benson2016higher,yin2017local}.
In~\cite{wang2014identification}, Wang\et proposed to measure the importance of a node in a network by its participation in different motifs. 
In~\cite{zhang2017structinf}, Zhang\et proposed to predict users' behaviors based on structural influence, i.e., the influence from a specific structure he/she appears. Compared to these previous studies, we utilize motif to explore the higher-order relations in pairs of nodes and incorporate them into the node authority computation with PageRank. In other words, we consider motifs from a global perspective while previous works only consider local motif structures of the nodes.

\section{Motif-based PageRank}
\label{sec-framework}
In this section, we introduce in detail our framework and algorithm. 

\subsection{Problem Formulation}
PageRank~\cite{page1999pagerank} was initially proposed to compute the authority of a webpage given the webgraph. Generally speaking, the authority of a node $p$ (webpage) depends on the number of incoming edges (hyperlinks) and the authority of nodes $q$ pointing to $p$. Let $\mG=(\mU, \mE, \bW)$, where $\mU = \{u_i|i = 1,,...,N\}$ is the node set, and $\mE = \{e_{ij}|i,j = 1,...,N\}$ is the edge set, where $e_{ij}$ is an edge from $u_i$ to $u_j$. According to~\cite{page1999pagerank}, the PageRank value $x_{u_i}$ of a node $u_i$ is defined in an iterative manner as:
\begin{equation}
x_{u_i} = d \cdot \sum_{u_j \in F_-(u_i)} \frac{x_{u_j}}{|F_+(u_j)|} + (1 - d),
\end{equation}
where $F_-(u_i)$ is the set of nodes pointing to $u_i$, and $F_+(u_j)$ is the set of nodes pointed by $u_j$. $d \in (0,1)$ is a damping factor, which assigns some weights to nodes without incoming edges.  When stacking all PageRank values into a vector $\bx$, the equation becomes
\begin{equation}
\label{eq-pagerank}
\bx = d\cdot \bP^T\bx + \frac{1 - d}{N}\be,
\end{equation}
where $\bx \in \mathbb{R}^N$ and $\bx_i$ is the PageRank value of the $i$-th node in $\mG$, and $N$ is the number of nodes. $\be \in \mathbb{R}^N$ is a vector with every entry equal to $1$. $\bP$ is the transition probability matrix obtained by $\bP_{ij}=\bW_{ij}/\sum_j\bW_{ij}$, where $\bW$ is the adjacency matrix of the graph and $\bW_{ij}$ represents the weight of $e_{ij}$. For a directed unweighted graph, $\bW_{ij} = 1$ if $e_{ij}$ exists and $\bW_{ij} = 0$ otherwise. 
In~\cite{page1999pagerank}, a simple iterative algorithm is used to compute the PageRank vector
\begin{equation}
\bx_t = d\cdot \bP^T\bx_{t-1} + \frac{1 - d}{N}\be,
\end{equation}
where $x_t$ is the PageRank vector in step $t$. In~\cite{bianchini2005inside}, Bianchini\et proved that this iterative computation always converges, thus we obtain the PageRank value for each node in the graph. 

Given a social network $\mG$, the PageRank value of each user represents his/her influence, and  $\bW_{ij}$ represents the existence or strength of friendship between user $u_i$ and $u_j$. From Eq.~\eqref{eq-pagerank}, we can see that $\bW$ affects the transition probability matrix $\bP$, thus the final PageRank values. In previous works~\cite{xing2004weighted,radicchi2009diffusion,zyczkowski2010citation,ding2011applying}, weighted PageRank has been shown to be effective for ranking researchers in citation networks. However, no matter original or weighted PageRank, the weights are calculated based on the direct relations between two researchers, e.g., the frequency of a researcher citing another one, or assigning more weights to more influential researchers. In other words, they only consider the first-order relations between two nodes, while ignoring higher-order relations captured by motifs. As shown in Figure~\ref{fig-motif-example}(a),  assuming it is a unweighted directed graph, if we consider directed edges only, $\bW_{12}$ and $\bW_{14}$ have the same weights because $u_2, u_3$ and $u_4$ have the same influence on $u_1$. However, if we also consider the triangular structure $u_1$ and $u_2$ participate in, it is clear that the $u_2$ has higher influence on $u_1$ than $u_4$. Therefore, $\bW_{12}$ should be larger than $\bW_{14}$. Therefore, it is desirable if the weight can incorporate information about the higher-order relations.

%Given a social network $\mG$, we can generalize this definition as follows. If there is an edge $e_{ij}$ from node $u_i$ to $u_j$, then we use $\bW_{ij}$ to represent the strength of the endorsement $u_i$ gives to $u_j$ or, conversely, the strength of the influence $u_j$ exerts on $u_i$. 
%The weights of endorsements user $u_i$ gives to all of his/her out-going neighbors are set to the same value, typically normalized to $\frac{1}{N_+(u_i)}$, where $N_+(u_i)$ represent the neighbors pointed from $u_i$, i.e., all of the users endorsed by $u_i$. As shown in Figure~\ref{fig-motif-example}(a),  if we consider directed edges only, $\bW_{12}$ and $\bW_{14}$ have the same weights because $u_2, u_3$ and $u_4$ have the same influence on $u_1$. However, if we also consider the triangular structure $u_1$ and $u_2$ participate in, it is clear that the $u_2$ has higher influence on $u_1$ than $u_4$. Therefore, $\bW_{12}$ should be larger than $\bW_{14}$. Therefore, it is desirable if the weight can incorporate information about the higher-order relations.

In the rest of this section, we first introduce the formal definition of motif for characterizing higher-order relations. Then, we give the definition of motif-based adjacency matrix and the computation methods for different types of motifs, including 3-node simple and anchor motifs, 4-node and 5-node motifs. Finally, we describe the MPR framework, which combines the edge-based and motif-based adjacency matrices in linear and non-linear ways.

\begin{table*}[t]
	\centering
	\caption{Computation of motif-based adjacency matrices for $M_1$ to $M_7$ in Figure~\ref{fig-motif-example}(b). The computation is given in~\cite{bensonsupplementary}.}
	\label{tb-motif-adj-computation}
	\begin{tabular}[\columnwidth]{clc}
		\toprule
		Motif & Matrix Computation &$\bW_{M_i} = $ \\ \midrule
		$M_1$   & $\bC = (\bU \cdot \bU) \odot \bU^T$  & $\bC + \bC^T$  \\
		$M_2$   & $\bC = (\bB \cdot \bU) \odot \bU^T + (\bU \cdot \bB) \odot \bU^T + (\bU \cdot \bU) \odot \bB$   & $\bC + \bC^T$  \\ 
		$M_3$   & $\bC = (\bB \cdot \bB) \odot \bU + (\bB \cdot \bU) \odot \bB + (\bU \cdot \bB) \odot \bB$   & $\bC + \bC^T$  \\
		$M_4$   & $\bC = (\bB \cdot \bB) \odot \bB$  & $\bC$  \\ 
		$M_5$   & $\bC = (\bU \cdot \bU) \odot \bU + (\bU \cdot \bU^T) \odot \bU + (\bU^T \cdot \bU) \odot \bU$   & $\bC + \bC^T$  \\
		$M_6$   & $\bC = (\bU \cdot \bB) \odot \bU + (\bB \cdot \bU^T) \odot \bU^T + (\bU^T \cdot \bU) \odot \bB$   & $\bC$  \\
		$M_7$   & $\bC = (\bU^T \cdot \bB) \odot \bU^T + (\bB \cdot \bU) \odot \bU + (\bU \cdot \bU^T) \odot \bB$   & $\bC$  \\
		\bottomrule
	\end{tabular}
\end{table*}

\begin{figure}[t]
	\centering
	\includegraphics[width=0.8\columnwidth]{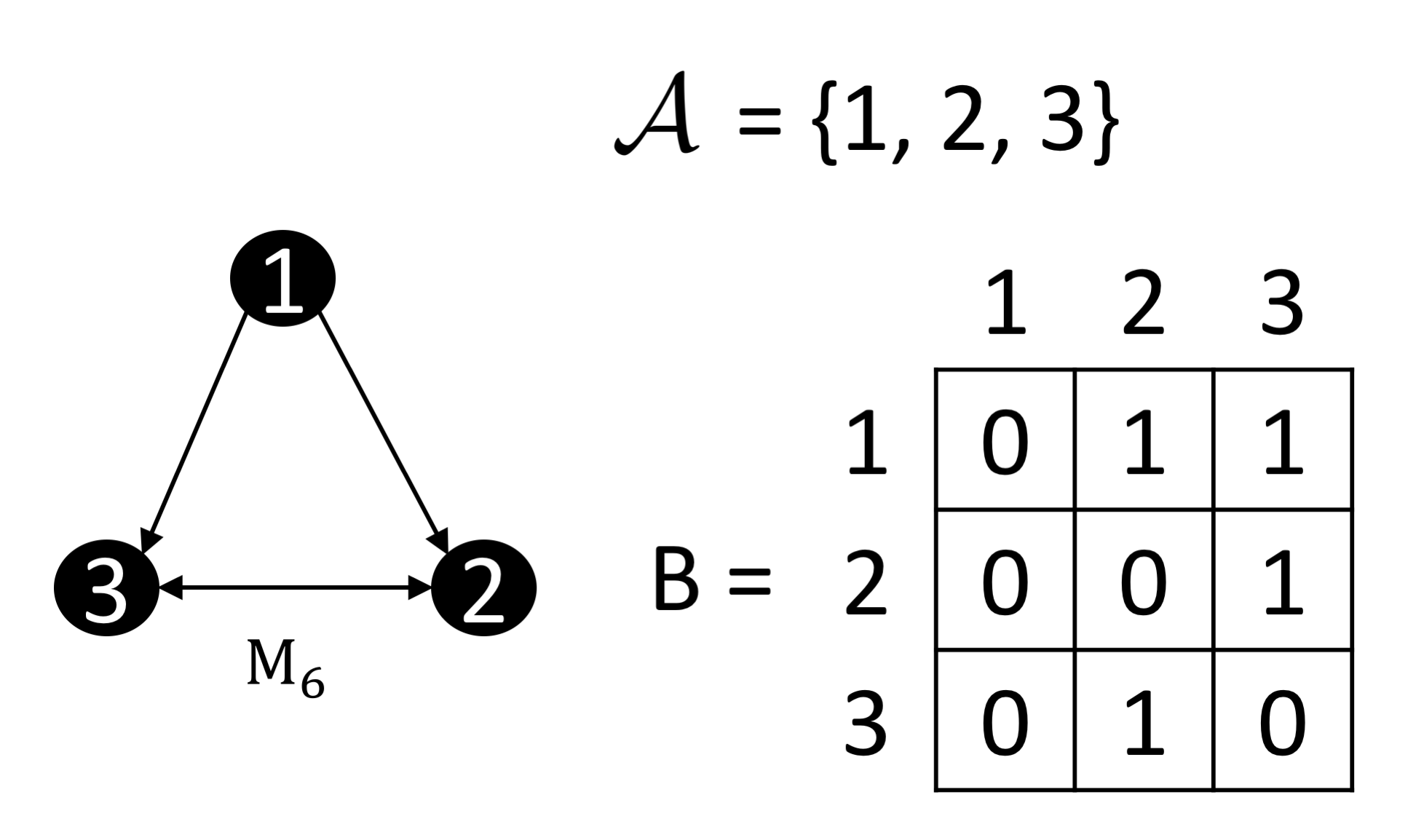}
	\caption{An example motif $M_6$ from Figure~\ref{fig-motif-example}(b). $M_6$ is a simple motif since the anchor set $\mA$ includes all of the nodes in $M_6$. $\bB$ is the binary matrix recording the edge pattern of $M_6$.}
	\label{fig-motif-def}
\end{figure}

\subsection{Motif Definitions}
We adopt the definitions of network motif and motif set from~\cite{bensonsupplementary}.
\begin{defn}
	\label{def-motif}
	\textbf{Network Motif.} A motif $M$ is defined on $k$ nodes by a tuple $(\bB, \mA)$, where $\bB$ is a $k \times k$ binary matrix, and $\mA \subset \{1,2,...,k\}$ specifies the \textit{anchor set}, which is the set of the indices of the \textit{anchor nodes}.
\end{defn}

Essentially, motif can be regarded as a pattern of edges on a small number of nodes from a graph. Here $\bB$ represents a graph encoding the edge patterns between the $k$ nodes, and $\mA$ denotes a subset of the $k$ nodes for defining the motif-based adjacency matrix. In other words, two nodes will be regarded as occurring in a given motif only when their indices belong to $\mA$.  Usually, anchor nodes are all of the $k$ nodes. In this case, the motif is called \textit{simple motif}; otherwise, it is called \textit{anchored motif}. In Figure~\ref{fig-motif-example}(b), we show 7 simple motifs, while in Figure~\ref{fig-anchor-motif-example}, we show 13 anchor motifs.

Given a motif definition, we can define the set of motif instances as follows.
\begin{defn}
	\textbf{Motif Set.} The motif set, denoted as $\mM(\bB,\mA)$, is defined in an unweighted directed graph $\mG$ with an adjacency matrix $\bW$ as:
	\begin{align}
	\mM(\bB,\mA) = \{(\text{set}(\bv),\text{set}(\chi_\mA(\bv))) | \bv \in V^k, \nonumber\\ \bv_1,...,\bv_k, \text{distinct},\nonumber \bW_{\bv} = \bB\}.
	\end{align}
\end{defn}
\noindent
where $\bv$ is an ordered vector representing the indices of $k$ nodes, and $\bW_{\bv}$ is the $k \times k$ adjacency matrix of the subgraph induced by $\bv$. $\chi_\mA$ is a selection function that takes the subset of a $k$-tuple indexed by $\mA$, and set($\cdot$) is an operator that transforms an ordered tuple to an unordered set, $\text{set}((\bv_1, \bv_2,...,\bv_k)) = \{\bv_1,\bv_2,...,\bv_k\}$.

The set operator is used to avoid duplicates when $\mM(\bB,\mA)$ is defined for motifs exhibiting symmetries. Therefore, we will just use $(\bv,\chi_\mA(\bv))$ to denote $(\text{set}(\bv), \text{set}(\chi_\mA(\bv)))$ when we discuss elements of $\mM(\bB,\mA)$. When $\bB$ and $\mA$ are arbitrary or clear from the context, we will simply denote the motif set by $\mM$.
Then, any $(\bv, \chi_\mA(\bv)) \in \mM$ is called a \textit{motif instance}.

\subsection{Motif-based Adjacency Matrix}
In this work, we propose to use motif to capture the higher-order relations between the nodes in a graph. Following the definition in~\cite{bensonsupplementary}, when given a motif set $\mM(\bB,\mA)$, we use the co-occurrence of two nodes in the anchor set to capture the corresponding higher-order relations. 
The motif-based adjacency matrix or co-occurrence matrix of a motif $\mM$ is defined as:
\begin{equation}
(\bW_M)_{ij} = \sum\limits_{(\bv,\chi_\mA(\bv)) \in \mM} \b1(\{i,j\} \subset \chi_\mA(\bv)),
\end{equation}
where $i \neq j$, and $\b1(s)$ is the truth-value indicator function, i.e., $\b1(s) = 1$ if the statement $s$ is true and 0 otherwise. Note that the weight is added to $(\bW_M)_{ij}$ only if $i$ and $j$ appear in the set of indices of anchor nodes. A concrete example for this definition is given in Figure~\ref{fig-motif-adj-computation-example}.

The motif-based adjacency matrix represents the frequency of two nodes appearing in a given motif. The larger $(\bW_M)_{ij}$ is, the more significant the relation between $i$ and $j$ is within the motif.
Then given a motif $M_k$, if we want to capture the higher-order relations, we need to construct the motif-based adjacency matrix $\bW_{M_k}$. 
The procedure is related to subgraph counting in large graphs, which has been extensively explored in the literature~\cite{ahmed2015efficient,jha2015path,han2016waddling,pinar2017escape}. We explore motifs with different sizes, including 3-node, 4-node and 5-node motifs. For 3-node motifs, we only consider the triangular ones because of the importance of triadic closure in social networks~\cite{simmel1908sociology,granovetter1977strength}. Further,
triangular motif-based adjacency matrices can be computed based on simple matrix computation, but there are no matrix computation methods available for 4-node and 5-node motifs~\cite{bensonsupplementary}.
Since it is expensive to enumerate all of the corresponding subgraphs, we derive an approximation algorithm to compute the 4-node and 5-node motif-based matrices based on a sampling method~\cite{kashtan2004efficient}.

\subsubsection{3-node simple motifs}
\begin{figure*}
	\centering
	\includegraphics[width=0.7\textwidth]{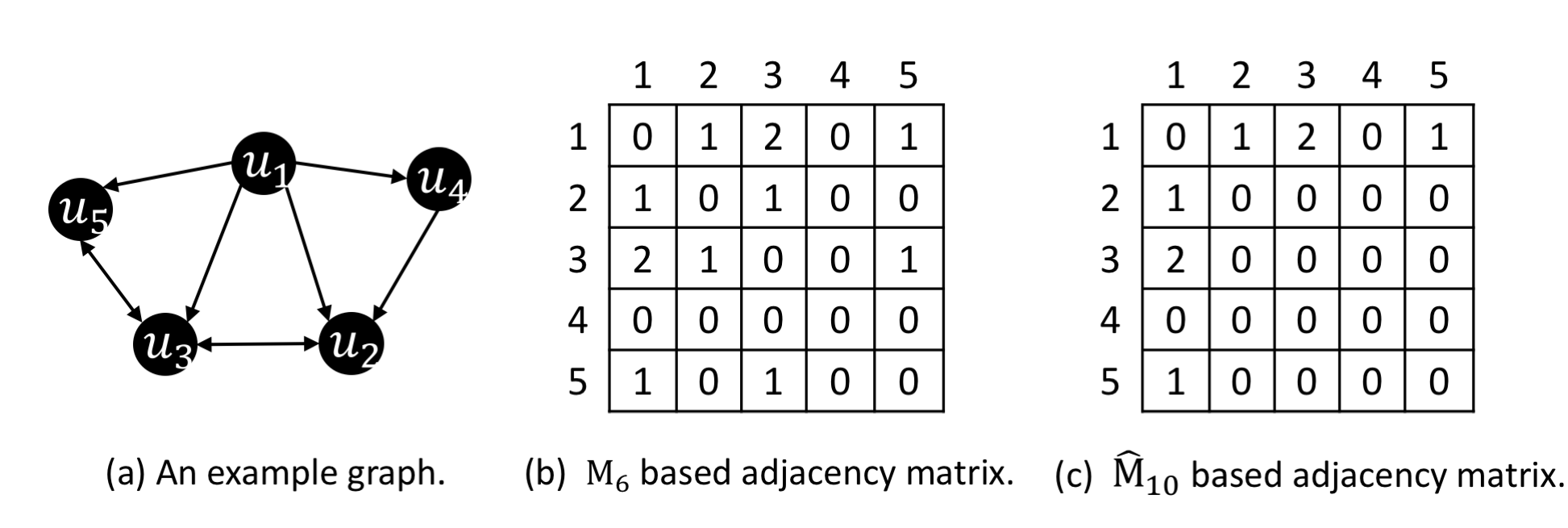}
%	\subfigure[An example graph with five nodes.] {\includegraphics[width=0.25\textwidth]{example-graph}}
%	\subfigure[$M_6$-based adjacency matrix.] {\includegraphics[width=0.25\textwidth]{m6_adj}}
%	\subfigure[$\hat{M}_{10}$-based adjacency matrix.] {\includegraphics[width=0.25\textwidth]{hat_m10_adj}}
	\caption{An example for computing $M_6$ and $\hat{M}_{10}$ based adjacency matrices according to the equation in Table~\ref{tb-motif-adj-computation} and \ref{tb-anchor-motif-adj-computation}, respectively. For example, $(\bW_{M_6})_{13} = (\bW_{\hat{M}_{10}})_{13} = 2$ because node $u_1$ and $u_3$ appear in two instances of $M_6$ and $\hat{M}_{10}$, i.e., $\{u_1,u_2,u_3\}$ and $\{u_1,u_3,u_5\}$. However,$(\bW_{M_6})_{35} = 1$, while $(\bW_{\hat{M}_{10}})_{35} = 0$ because of the different set of anchor nodes. Although nodes $u_3$ and $u_5$ occur in the subgraph corresponding to $\hat{M}_{10}$, they do not occur in the anchor set.}
	\label{fig-motif-adj-computation-example}
\end{figure*}

In this part, we show how to compute motif-based adjacency matrices for 3-node simple motifs. Let $\bW$ be the adjacency matrix for $\mG$, and $\bU$ and $\bB$, respectively, be the adjacency matrix of the unidirectional and bidirectional links of $\mG$. As in~\cite{benson2016higher}, we focus on unweighted graphs where elements in $\bW$ are either ones or zeros. For example, in Figure~\ref{fig-motif-adj-computation-example}(a), $e_{23}$ connecting $u_2$ and $u_3$ is a bidirectional edge while $e_{12}$ connecting $u_1$ and $u_2$ is unidirectional. Then we have $\bB = \bW \odot \bW^T$ and $\bU = \bW - \bB$, where $\odot$ denotes the Hadamard (entry-wise) product. Note that $\bB$ is a binary matrix representing the existence of bidirectional links between two nodes in a directed graph. The computation of the adjacency matrices of all seven 3-node motifs is summarized in Table~\ref{tb-motif-adj-computation}. Note that the equations in Table~\ref{tb-motif-adj-computation} are given in~\cite{bensonsupplementary}, and we only select those for the seven triangular motifs in Figure~\ref{fig-motif-example}(b).

We use $M_6$ in Figure~\ref{fig-motif-def} to illustrate the computation of the motif-based adjacency matrix. Taking two arbitrary nodes, $u_i$ and $u_j$, we use $(\bW_{M_6})_{ij}$ to record the frequency of their participation in $M_6$.
There are six different cases of $u_i$ and $u_j$ occurring in $M_6$.
For clarity, we use $1,2,3$ to denote the positions where a node can occur. Then, $u_i$ and $u_j$ intermediated with $u_k$ can generate six position combinations, i.e., $\{(3,1,2),(2,1,3)\},\{(1,2,3),(3,2,1)\}, \{(1,3,2),(2,3,1)\}$, 
where we assume the three nodes to be ordered are $(u_i, u_j, u_k)$. 
As shown in Figure~\ref{fig-motif-def}, $e_{23}$ is a bidirectional edge, while $e_{13}$ and $e_{12}$ are unidirectional edges. To compute the frequency of $u_i$ and $u_j$ participations in $M_6$, we need to add up their frequencies in all six cases. 
In the case $(3,1,2)$, i.e., $u_i$ is in position 3 and $u_j$ is in position 2, the frequency can be obtained by $(\bU \cdot \bB) \odot \bU$, where $\bU \cdot \bB$ is for path $3 \rightarrow 1 \leftrightarrow 2$ without edge $e_{32}$, and $\odot \bU$ just complements the motif with edge $e_{32}$. 
In this way, we get the frequency of $u_i$ and $u_j$ appearing in a specific case of $M_6$, i.e., $(3,1,2)$. Similarly, we can obtain the frequency of the other cases. Note that, due to the symmetry of $(3,2,1)$ and $(3,1,2)$, we only need to add up three equations as shown in Table~\ref{tb-motif-adj-computation}. 
Figure~\ref{fig-motif-adj-computation-example}(b) shows the motif-based adjacency matrix for $M_6$.

\begin{figure*}
	\centering
	\includegraphics[width=0.9\textwidth]{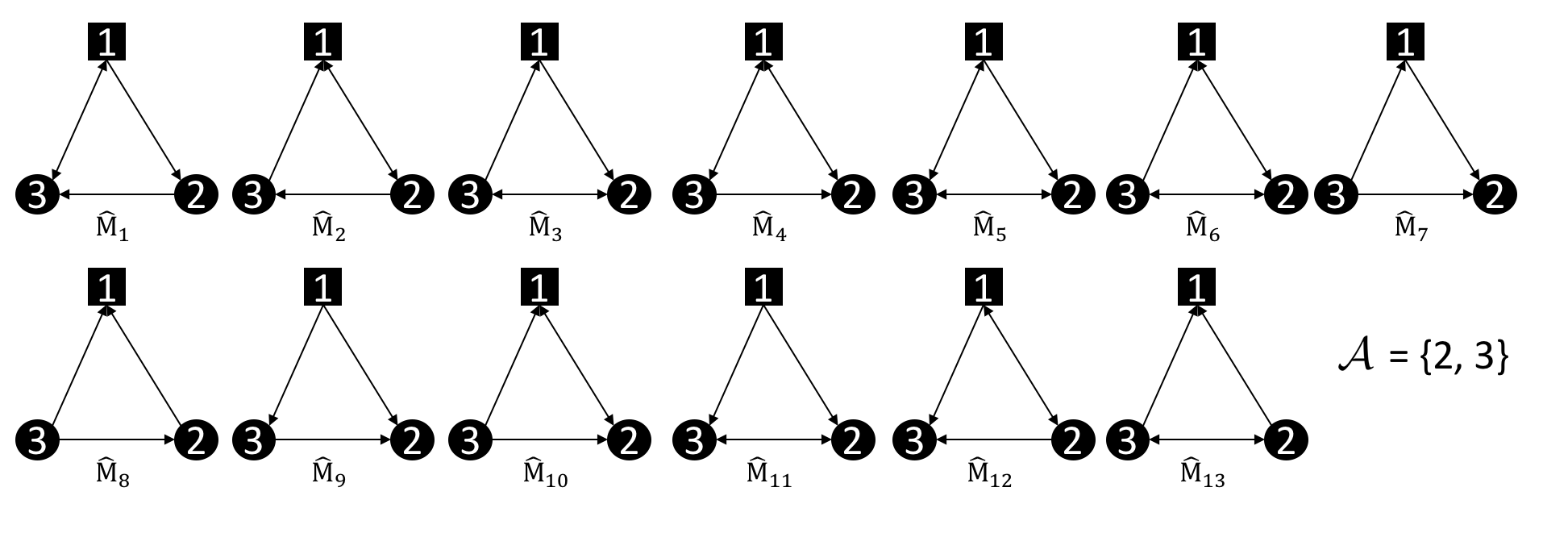}
	\caption{All triangular anchor motifs in a directed unweighted graph. Note that here $(\bW_M)_{ij}$ are representing frequency of $i$ and $j$ participating in the anchor set of motif, i.e., $\mA = \{2,3\}$. It means $(\bW_M)_{ij}$ will be added by one only if $i$ and $j$ occur in the positions 2 and 3.}
	\label{fig-anchor-motif-example}  
\end{figure*}

\begin{table*}[t]
	\centering
	\caption{Computation of motif-based adjacency matrices for $\hat{M}_1$ to $\hat{M}_{13}$ in Figure~\ref{fig-anchor-motif-example}.}
	\label{tb-anchor-motif-adj-computation}
	\begin{tabular}{clc|clc}
		\toprule
		Motif & Matrix Computation &$\bW_{\hat{M}_i} = $ & Motif & Matrix Computation &$\bW_{\hat{M}_i} = $\\ \midrule
		$\hat{M}_1$  &$\bC = (\bB \cdot \bU) \odot \bU^T$  & $\bC + \bC^T$ &$\hat{M}_8$      & $\bC = (\bU \cdot \bU^T) \odot \bU$ & $\bC + \bC^T$ \\		
		$\hat{M}_2$  & $\bC = (\bU \cdot \bB) \odot \bU^T$  & $\bC + \bC^T$& $\hat{M}_{9}$   & $\bC = (\bU^T \cdot \bU) \odot \bU$ & $\bC + \bC^T$ \\
		$\hat{M}_3$  & $\bC = (\bU \cdot \bU) \odot \bB$   & $\bC + \bC^T$ & $\hat{M}_{10}$   & $\bC = (\bU \cdot \bB) \odot \bU$   & $\bC + \bC^T$ \\
		$\hat{M}_4$  & $\bC = (\bB \cdot \bB) \odot \bU$  & $\bC + \bC^T$&   $\hat{M}_{11}$   & $\bC = (\bU^T \cdot \bU) \odot \bB$ & $\bC$ \\
		$\hat{M}_5$  & $\bC = (\bB \cdot \bU) \odot \bB$  & $\bC + \bC^T$&   $\hat{M}_{12}$   & $\bC = (\bU^T \cdot \bB) \odot \bU^T$& $\bC + \bC^T$\\	
		$\hat{M}_6$  & $\bC = (\bU \cdot \bB) \odot \bB$ & $\bC + \bC^T$ &   $\hat{M}_{13}$   & $\bC = (\bU \cdot \bU^T) \odot \bB$  & $\bC$  \\		
		$\hat{M}_7$   & $\bC = (\bU \cdot \bU) \odot \bU$ & $\bC + \bC^T$&    \\
		\bottomrule
	\end{tabular}
\end{table*}

\subsubsection{3-node anchor motifs}
As Definition~\ref{def-motif} indicates, the difference between simple and anchor motifs is the size of their anchor sets. In Figure~\ref{fig-anchor-motif-example}, we show all triangular anchor motifs, denoted as $\hat{M}_1, ..., \hat{M}_{13}$, where circular nodes are anchor nodes.
For example, $\hat{M}_{10}$ is ``anchored'' by the square node, which means
the motif-based adjacency matrix will add one to the weight of $e_{ij}$ when two nodes are connected by a third intermediate node. $\hat{M}_{10}$ can be regarded as one of the relations belonging to the simple motif $M_6$. Specifically, $\hat{M}_{10}$ represents the example case discussed in the previous section, i.e., when $(u_i, u_j, u_k)$ are in the positions $(3,1,2)$ or $(1,3,2)$. Therefore, the $\hat{M}_{10}$-based adjacency matrix $(\bW_{\hat{M}_{10}})_{ij}$ can be computed by $(\bU \cdot \bB) \odot \bU + (\bB \cdot \bU^T) \odot \bU^T$, and the result is shown in Figure~\ref{fig-motif-adj-computation-example}(c).
In conclusion, the motif-based adjacency matrices of all triangular anchor motifs can be computed in a similar way. Equations for the matrix computation is given in Table~\ref{tb-anchor-motif-adj-computation}.

\subsubsection{4-node and 5-node motifs}
\label{sec-4-5-motif}
According to~\cite{bensonsupplementary}, there are no simple matrix computational methods for 4-node and 5-node  motif-based adjacency matrices. Thus, we need to count all of the subgraphs to identify the number of a specific motifs. In this section, we depict how we use subgraph counting methods to compute the motif-based adjacency matrices.

There are two challenges when the size of motifs becomes large. First, the computation cost increases dramatically. Second, there will be too many isomorphic types of subgraphs with a given number of nodes. For example, there are 13 different 3-node subgraphs, 199 different 4-node subgraphs, and 9364 different 5-node subgraphs~\cite{kashtan2004efficient}. However, not all subgraphs can be regarded as motifs. As defined in~\cite{milo2002network}, motifs are those subgraphs which occur statistically significantly more often in real networks than in a randomized network. In practice, we need to construct a number of randomized networks to identify motifs from all the subgraphs. Therefore, the computational cost will further increase with a large number of isomorphic subgraphs.

In the literature, many methods have been proposed to estimate the number of subgraphs of different sizes \cite{ahmed2015efficient,jha2015path,han2016waddling,pinar2017escape}, which are all for undirected graphs. However, the work reported in this paper not only deals with directed graph, which leads to more types of motifs, but also considers the pairwise relations between nodes. To address these unique challenges, we adopt a sampling method in~\cite{kashtan2004efficient}, which shows good performance in estimating the ratio of the number of different n-node motifs. Given the size of subgraph $n$, the method can sample a number of n-node subgraphs according to the probability distribution of the frequency of different types of n-node subgraphs. Based on the sampled subgraphs, we derive our algorithm to compute the motif-based adjacency matrices for different types of motifs. 

The detailed sampling algorithm is given in~\cite{kashtan2004efficient}. We use it to sample a set of n-node subgraphs. Then, for each motif, we obtain the set of all motif instances, denoted as $S_{M_k}$.
The detailed algorithm for constructing n-node motif-based adjacency matrices is
given in Algorithm~\ref{alg-motif-4-5-adj-computation}. Note that at Line 4, the occurrence of nodes $i$ and $j$ will count as one when they are directly connected. 

Note that we omit some technical details for clarity. In the experimental section, we give more details about how to compute motif-based adjacency matrices for 4-node and 5-node motifs.

\begin{algorithm}[ht]
	\caption{Sampling-based Motif-based Matrix Computation.}
	\begin{algorithmic}[1]
		\REQUIRE {The set of sampling instances of $S_{M_k}$, the adjacency matrix $\bW$;}
		\FOR{$(\bv, \chi_\mA(\bv)) \in S_{M_k}$}
		\FOR{$i \in \bv$}
		\FOR{$j \in \bv, j\neq i$}
		\IF {$\bW_{ij} > 0$}
			\STATE{$(\bW_{M_k})_{ij}$++;}
		\ENDIF
		\ENDFOR
		\ENDFOR
		\ENDFOR
		\RETURN{$\bW_{M_k}$}
	\end{algorithmic}
	\label{alg-motif-4-5-adj-computation}
\end{algorithm}

\subsubsection{Computation Analysis}

In this part, we analyze the computational complexity for 3-node simple and anchor motifs, 4-node, and 5-node simple motifs.

In the formulas for computing $\bW_{M_i}$, displayed in Table~\ref{tb-motif-adj-computation}, the core computation kernel is $(\bX \cdot \bY) \odot \bZ$, which can be efficiently computed because the matrices are sparse. The computational cost is proportional to the numbers of columns and rows, as well as the number of non-zero elements in the sparse matrices. The computation of motif-based adjacency matrices for 3-node anchor motifs is shown in Table~\ref{tb-anchor-motif-adj-computation}.  

When computing the 4-node and 5-node motif-based matrices, we first execute the sampling algorithm, which has total computation complexity of $O( N \times n^{n+1})$, where $n$ is the size of the motif and $N$ is the number of samples~\cite{kashtan2004efficient}.
The major cost of Algorithm~\ref{alg-motif-4-5-adj-computation} is incurred in the three \textit{for} loops, which has complexity $O(|S_{M_k}| \times n^2)$. Therefore, the overall complexity is $O( N \times n^{n+1}) + O(|S_{M_k}| \times n^2)$, i.e., $O( N \times n^{n+1})$. Thus, the computational cost for computing 4-node and 5-node motifs is feasible for large scale social networks, as demonstrated in our experiments.

\subsection{Motif-based PageRank}
\label{sec-higher-order-pg}
In this paper, we argue that the edge-based and motif-based relations are complementary to each other in user ranking. Thus, we study a linear and a non-linear method to effectively combine the edge-based and motif-based adjacency matrices.
%In this way, motif-based edges can be regarded as supplementary to conventional authority computation with PageRank.
%
%We propose to use a linear and non-linear combinations to fuse the edge-based and motif-based adjacency matrices.

\textbf{Linear Combination.} For a given motif $M_k$, the linear combination method is defined as follows:
\begin{equation}
\label{eq-linear-combination}
\bH_{M_k} = \alpha \cdot \bW + (1 - \alpha) \cdot \bW_{M_k}.
\end{equation}

\textbf{Non-Linear combination.} For a given motif $M_k$, the non-linear combination method is defined as follows:
\begin{equation}
\label{eq-nonlinear-combination}
\bH_{M_k} = \bW^{\alpha} \odot \bW_{M_k}^{1 - \alpha}.
\end{equation}
In Eq.~\eqref{eq-linear-combination} and \eqref{eq-nonlinear-combination}, $\alpha \in [0, 1]$, which balances the edge-based and motif-based relations.

In our proposed MPR framework, we run the PageRank algorithm on the combined adjacency matrix. Specifically, we compute the transition probability matrix $({\bP_{M_k}})_{ij}=({\bH_{M_k}})_{ij}/\sum_j({\bH_{M_k}})_{ij}$ and substitute ${\bP_{M_k}}$ for the transition probability matrix $\bP$ in Eq.~(\ref{eq-pagerank}).

\section{Experiments}
\label{sec-exp}
In this section, we describe the details of the experiments,
and compare the performance of MPR to several baselines, including the
original (unweighted) and weighted PageRank algorithms, to demonstrate the effectiveness of MPR for user ranking in social networks.

\subsection{Datasets and Settings}
\label{subsec-unsup-exp-settings}
Our experiments are conducted on three real-world social datasets. The first is a scholar network, DBLP, which is provided by ArnetMiner~\cite{tang2008arnetminer}. The other two are trust networks, Epinions and Ciao, which are provided by~\cite{tang-etal12a} and \cite{tang-etal12b}, respectively. More information about the datasets is given below.

\noindent\textbf{DBLP.} We use the DBLP dataset (version V8) in the AMiner Website.\footnote{https://cn.aminer.org/billboard/citation} DBLP is an academic dataset containing records of publications and their authors.
We focus on publication records in six research domains: ``Artificial Intelligence,'' ``Computer Vision,'' ``Database,'' ``Data Mining,'' ``Information Retrieval,'' and ``Machine Learning.''
After extracting the authors and papers from these domains, the citation network is constructed to evaluate the social influence of the authors. When constructing the citation network, we add an edge from $u_i$ to $u_j$ if author $u_i$ cites at least one paper of author $u_j$.

\noindent\textbf{Epinions and Ciao.} These are two review websites where users can write reviews on products and cast \textit{review helpfulness rating}
on reviews written by other users. Moreover, users can add other users into their trust lists if they like the reviews written by these users. We can expect that if a user's reviews receive high helpfulness ratings, it is more likely for him to become a trusted user of other users. When constructing the trust network, an edge $e_{ij}$ is added when user $u_i$ trusts user $u_j$.

The statistics of the datasets are depicted in Table~\ref{tb-dataset-stat}.

\begin{table}[t]
	\centering
	\caption{Statistics of the three datasets: DBLP, Epinions, Ciao. The density is computed by $\frac{\#edges}{\#nodes \times (\#nodes - 1)}$.}
	\label{tb-dataset-stat}
	\begin{tabular}[\columnwidth]{c|ccc}
		\toprule
		& Nodes & Edges & Density(\%)\\ \midrule
		DBLP & 35,315   & 941,936  &  0.076\\
		Epinions & 18,089  & 355,217  & 0.109  \\
		Ciao & 2,342  &  57,544 &  1.049 \\ 
		\bottomrule
	\end{tabular}
\end{table}

\begin{table*}[t]
	\centering
	\caption{NDCGs of the linear combination method for top10, top50, top500 users on DBLP, Epinions, and Ciao datasets. The results of the baselines, anchor motifs, and simple motifs are listed in different blocks. The results of motifs which outperform all baselines are in boldface, and the best performance of anchor motifs and simple motifs in each column is further underlined.}
	\label{tb-ndcg-moitf-3-linear}
	\begin{tabular}{c|ccc|ccc|ccc}
		\hline
		& \multicolumn{3}{c|}{DBLP} & \multicolumn{3}{c|}{Epinions} & \multicolumn{3}{c}{Ciao} \\ \hline
		TopK& 10 & 50 & 500 & 10 & 50 & 500 & 10 & 50 & 500 \\ \hline
		IND & 0.9879 & 0.9639 & 0.9400 & 0.9476 & 0.9563 & 0.9343 & 0.9218 & 0.8651 & 0.9120 \\ 
		BET & 0.9796 & 0.9710 & 0.9559 & 0.9566 & 0.9559 & 0.9403 & 0.9421 & 0.8961 & 0.8911 \\ 
		CLO & 0.9875 & 0.9614 & 0.9285 & 0.9308 & 0.9346 & 0.9382 & 0.9021 & 0.9225 & 0.9251 \\ 
		BPR & 0.9464 & 0.9414 & 0.9527 & 0.9777 & 0.9543 & 0.9365 & 0.8332 & 0.8599 & 0.8932 \\ 
		WPR & 0.9154 & 0.8871 & 0.9350 & 0.9777 & 0.9543 & 0.9365 & 0.8332 & 0.8599 & 0.8932 \\\hline
		$\hat{M}_1$    & 0.9352 & 0.9107 & 0.9434 & \textbf{0.9806} & \textbf{0.9612} & 0.9389 & \textbf{0.9851} & \textbf{0.9339} & \textbf{0.9275} \\
		$\hat{M}_2$    & \textbf{0.9894} & \underline{\textbf{0.9791}} & \textbf{0.9615} & 0.9554 & \textbf{0.9573} & \textbf{0.9456} & \textbf{0.9626} & 0.9113 & \textbf{0.9276} \\
		$\hat{M}_3$    & 0.9833 & 0.9564 & \textbf{0.9618} & 0.9528 & \textbf{0.9593} & \textbf{0.9409} & \textbf{0.9624} & 0.9202 & \textbf{0.9269} \\
		$\hat{M}_4$    & \textbf{0.9925} & \textbf{0.9772} & \underline{\textbf{0.9636}} & \textbf{0.9835} & 0.9559 & 0.9372 & \textbf{0.9584} & 0.9122 & \textbf{0.9271} \\
		$\hat{M}_5$    & 0.9785 & 0.9593 & 0.9548 & \textbf{0.9828} & 0.9558 & 0.9391 & \textbf{0.9500} & 0.9111 & \textbf{0.9271} \\
		$\hat{M}_6$    & 0.9860 & 0.9381 & \textbf{0.9568} & \textbf{0.9828} & 0.9563 & \textbf{0.9407} & \textbf{0.9457} & \textbf{0.9301} & \textbf{0.9274} \\
		$\hat{M}_7$    & 0.9541 & 0.9326 & 0.9514 & 0.9551 & \textbf{0.9646} & 0.9382 & 0.9241 & 0.9222 & \textbf{0.9270} \\
		$\hat{M}_8$    & 0.9859 & \textbf{0.9738} & \textbf{0.9583} & 0.9495 & 0.9560 & \textbf{0.9445} & 0.9132 & 0.9132 & \textbf{0.9273} \\
		$\hat{M}_9$    & \textbf{0.9907} & 0.9640 & \textbf{0.9628} & 0.9619 & \underline{\textbf{0.9672}} & 0.9370 & \textbf{0.9847} & \underline{\textbf{0.9421}} & \textbf{0.9277} \\
		$\hat{M}_{10}$ & 0.9533 & 0.9247 & 0.9506 & \underline{\textbf{0.9935}} & \textbf{0.9631} & 0.9387 & 0.9197 & 0.9202 & \textbf{0.9273} \\
		$\hat{M}_{11}$ & 0.9645 & 0.9448 & 0.9499 & \textbf{0.9798} & \textbf{0.9567} & 0.9389 & \underline{\textbf{0.9907}} & \textbf{0.9259} & \textbf{0.9274} \\
		$\hat{M}_{12}$ & \underline{\textbf{0.9934}} & 0.9603 & \textbf{0.9598} & 0.9581 & \textbf{0.9635} & 0.9386 & 0.9274 & \textbf{0.9230} & \textbf{0.9272} \\
		$\hat{M}_{13}$ & 0.9879 & 0.9516 & 0.9406 & 0.9602 & \textbf{0.9577} & \underline{\textbf{0.9461}} & \textbf{0.9809} & 0.9185 &\underline{\textbf{0.9298}} \\ \hline
		$M_1$ & 0.9753 & 0.9590 & \textbf{0.9623} & 0.9777 & \underline{\textbf{0.9656}} & \textbf{0.9406} & \textbf{0.9802} & \textbf{0.9347} & \textbf{0.9392} \\ 
		$M_2$ & \textbf{0.9890} & 0.9424 & \textbf{0.9585} & 0.9777 & \textbf{0.9581} & \textbf{0.9417} & \underline{\textbf{0.9905}} & \textbf{0.9453} & \textbf{0.9401} \\ 
		$M_3$ & \textbf{0.9895} & 0.9508 & \textbf{0.9586} & \textbf{0.9788} & \textbf{0.9568} & 0.9378 & \textbf{0.9768} & \textbf{0.9576} & \textbf{0.9441} \\ 
		$M_4$ & 0.9809 & 0.9477 & 0.9528 & \textbf{0.9827} & 0.9557 & 0.9395 & \textbf{0.9719} & \textbf{0.9357} & \textbf{0.9401} \\ 
		$M_5$ & 0.9877 & 0.9513 & \textbf{0.9574} & 0.9777 & 0.9551 & \textbf{0.9454} & \textbf{0.9792} & \underline{\textbf{0.9792}} & \textbf{0.9401} \\ 
		$M_6$ & 0.9634 & 0.9525 & \textbf{0.9588} & \underline{\textbf{0.9957}} & \textbf{0.9596} & 0.9382 & \textbf{0.9459} & \textbf{0.9459} & \underline{\textbf{0.9427}} \\ 
		$M_7$ & \underline{\textbf{0.9920}} & \underline{\textbf{0.9766}} & \underline{\textbf{0.9640}} & \textbf{0.9780} & \textbf{0.9614} & \underline{\textbf{0.9442}} & \textbf{0.9514} & \textbf{0.9500} & \textbf{0.9418} \\ \hline
	\end{tabular}
\end{table*}

\begin{table*}[t]
	\centering
	\caption{Significance test for \textit{BRP v.s. MPR} and \textit{WPR v.s. MPR} on all three datasets. For the DBLP dataset, $M_7$ is chosen for MPR, whereas $M_4$ is chosen for MPR both for the Epinions and Ciao datasets.}
	\label{tb-sig-test}
	\begin{tabular}{c|ccc|ccc|ccc}
		\hline
		& \multicolumn{3}{c|}{DBLP} & \multicolumn{3}{c|}{Epinions} & \multicolumn{3}{c}{Ciao} \\ \hline
		TopK& 10 & 50 & 500 & 10 & 50 & 500 & 10 & 50 & 500 \\ \hline
		BPR & 0.9287 & 0.8894 & 0.9357 & 0.8725 & 0.8460 & 0.8856 & 0.8753 & 0.8746 & 0.8912 \\ 
		p-value & 1.05e-34 & 6.10e-44 & 2.13e-29 & 2.40e-22 & 1.77e-11 & 2.49e-41 & 2.31e-06 & 2.94e-09 & 9.24e-19 \\ \hline
		WPR & 0.9050 & 0.8904 & 0.9354 & 0.8751 & 0.8492 & 0.8859 & 0.8603 & 0.8758 & 0.8937 \\ 
		p-value & 5.56e-28 & 2.40e-45 & 1.12e-33 & 2.22e-20 & 4.57e-10 & 9.85e-42 & 4.14e-11 & 1.17e-09 & 3.72e-14 \\ \hline
		MPR & \textbf{0.9864} & \textbf{0.9380} & \textbf{0.9431} & \textbf{0.9313} & \textbf{0.8779} & \textbf{0.9160} & \textbf{0.9469} & \textbf{0.9123} & \textbf{0.9082} \\ \hline
	\end{tabular}
\end{table*}

\noindent\textbf{Evaluation Metrics.} To evaluate the effectiveness of MPR, we compare the quality of the topK users ranked by the different algorithms. Specifically, we extract $K$ users with the largest PageRank values and then compute the Normalized Discounted cumulative Gain (NDCG), which is a popular metric for ranking quality in information retrieval (IR)~\cite{jarvelin2002cumulated}.
For topK results, DCG@K is defined as:
\begin{equation}
DCG_K = \sum_{i=1}^{K} \frac{rel_i}{\log_2(i+1)},
\end{equation}
where $rel_i$ represents the relevance score of a document to a given query. $\log_2(i+1)$ is used to penalize an algorithm if it ranks highly relevant items in low positions. The normalized $DCG_K$ is computed as:
\begin{equation}
NDCG_K = \frac{DCG_K}{IDCG_K},
\label{eq-ndcg}
\end{equation}
where $IDCG_K$ is the ideal ranking for the results, i.e. the results are sorted according to their relevance scores. In this way, it measures how good a ranking list is compared to the ideal ranking.

In our experiments, for the DBLP dataset, we use the H-indexes of the authors as the relevance scores. In the research community, H-index~\cite{hirsch2005index} is a commonly used metric to measure the influence of an author in the research community. It considers both the quality and quantity of the author's published papers based on the citation network. The larger the H-index an author has, the higher is his/her influence. We crawled the H-index of all authors in our dataset from the AMiner website\footnote{https://aminer.org/} before July 2017. For Epinions and Ciao datasets, we use the average helpfulness rating of a user's reviews as the user's trustworthiness score, which means that the larger the helpfulness rating is, the higher the user's trustworthiness.

\noindent\textbf{Baselines.} We compare MPR with the following methods:
\begin{itemize}
	\item \textbf{IND}: It selects influential nodes based on the incoming degrees of the nodes, i.e., for DBLP authors whose works are cited by most other authors, and for Epinions and Ciao users who are trusted by most other users.
	\item \textbf{BET}: It selects influential nodes based on the \textit{betweenness score}, which is a centrality measure defined as the number of times a node acts as a bridge along the shortest path between two other nodes~\cite{freeman1977set}.
	\item \textbf{CLO}: It selects influential nodes based on the \textit{closeness score}. The closeness score of a node is a centrality measure that is the average length of the shortest paths between the node and all other nodes in a graph~\cite{sabidussi1966centrality}. 
	\item \textbf{BPR}: This method runs PageRank in a binary network, where the weights of all edges are set to 1. 
	\item \textbf{WPR}: This method runs PageRank in a weighted network, where the weight of $e_{ij}$ is the frequency of $u_i$'s citing the work of $u_j$ for DBLP whereas, on Epinions and Ciao, the weight is still set to one since there is no weighting information in these two datasets.
\end{itemize}
\vspace{-0.2in}

For MPR, we have various settings. First, we study 3-node motifs, including simple and anchor ones. Each motif is used separately to compute the motif-based adjacency matrix. Then we run MPR with linear and non-linear combination methods and report the influence of the control parameter $\alpha$ in both cases according to Eq.~\eqref{eq-linear-combination} and \eqref{eq-nonlinear-combination}, respectively. Further, we explore motifs with larger sizes (4-node and 5-node motifs). Finally, we report the performance of MPR with ensemble of the seven 3-node simple motifs in Figure~\ref{fig-motif-example}(b). Through these extensive experiments, we demonstrate the effectiveness of incorporating motif-based higher-order relations into conventional edge-based relations for user ranking in social networks.

\begin{figure*}[!ht]
	\subfigure[$M_7$ on DBLP]{	\includegraphics[width=0.33\textwidth]{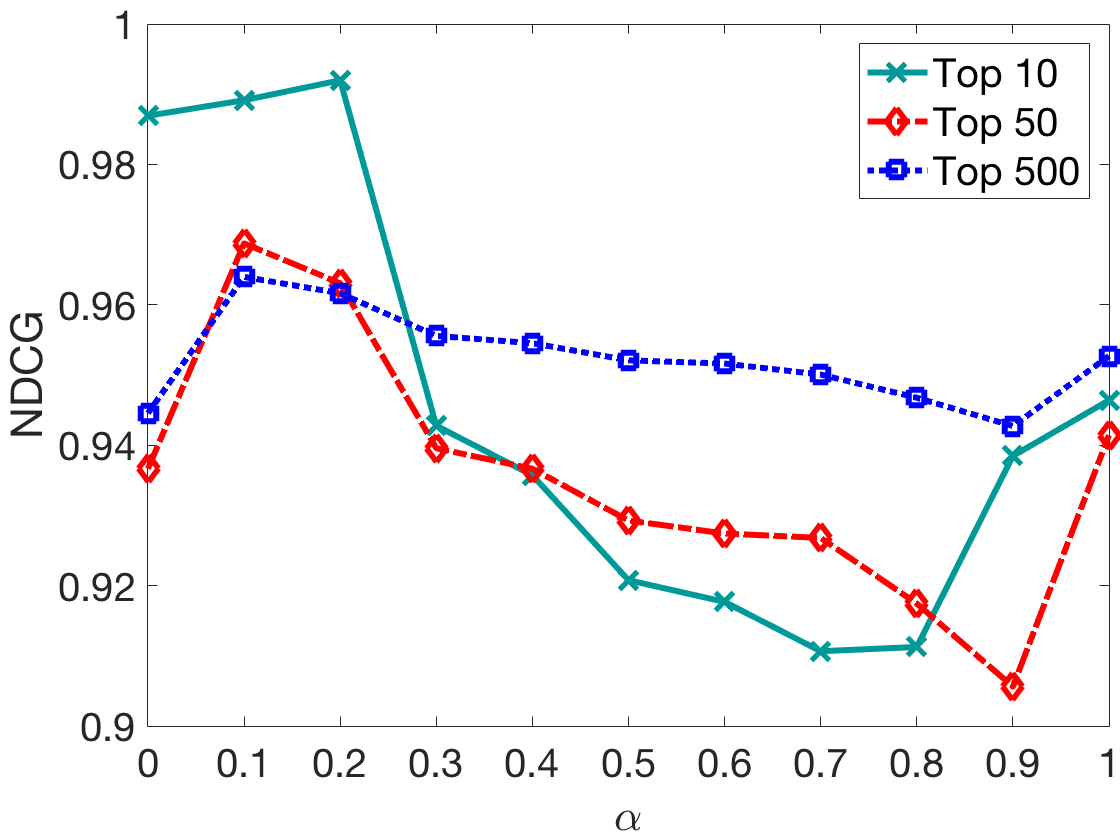}}
	\subfigure[$M_7$ on Epinions]{	\includegraphics[width=0.33\textwidth]{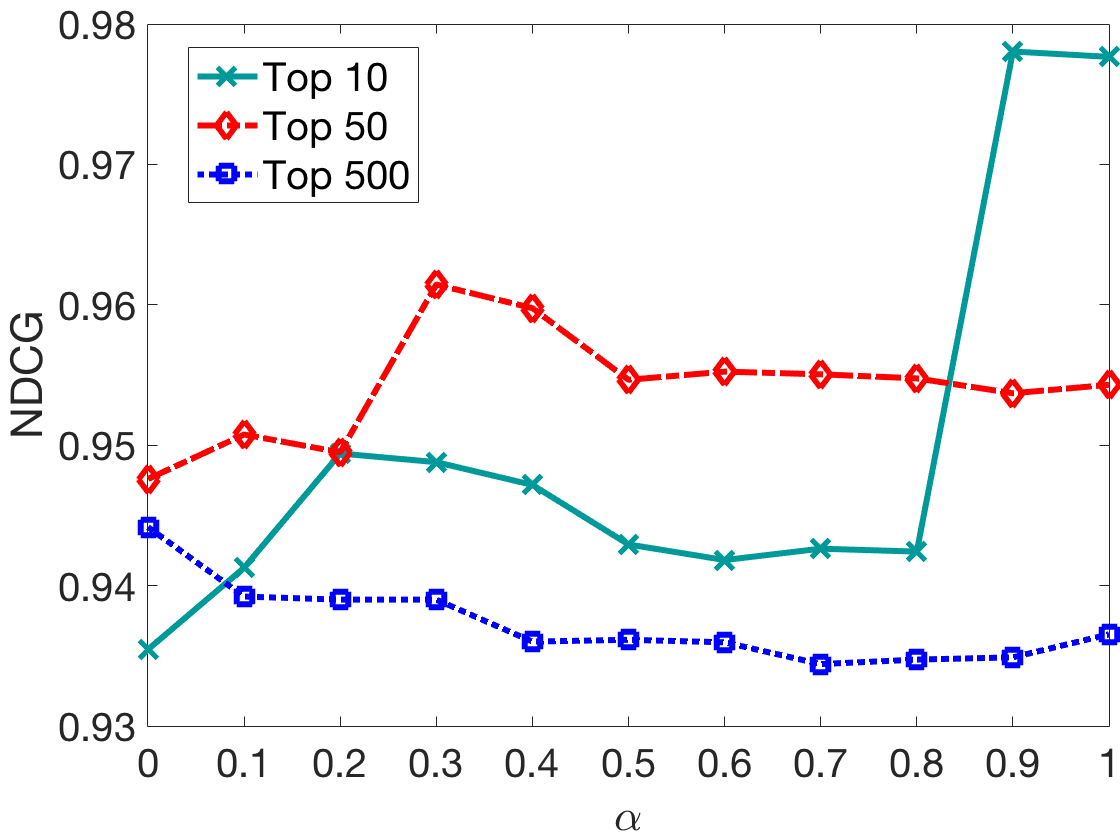}}
	\subfigure[$M_4$ on Ciao]{	\includegraphics[width=0.33\textwidth]{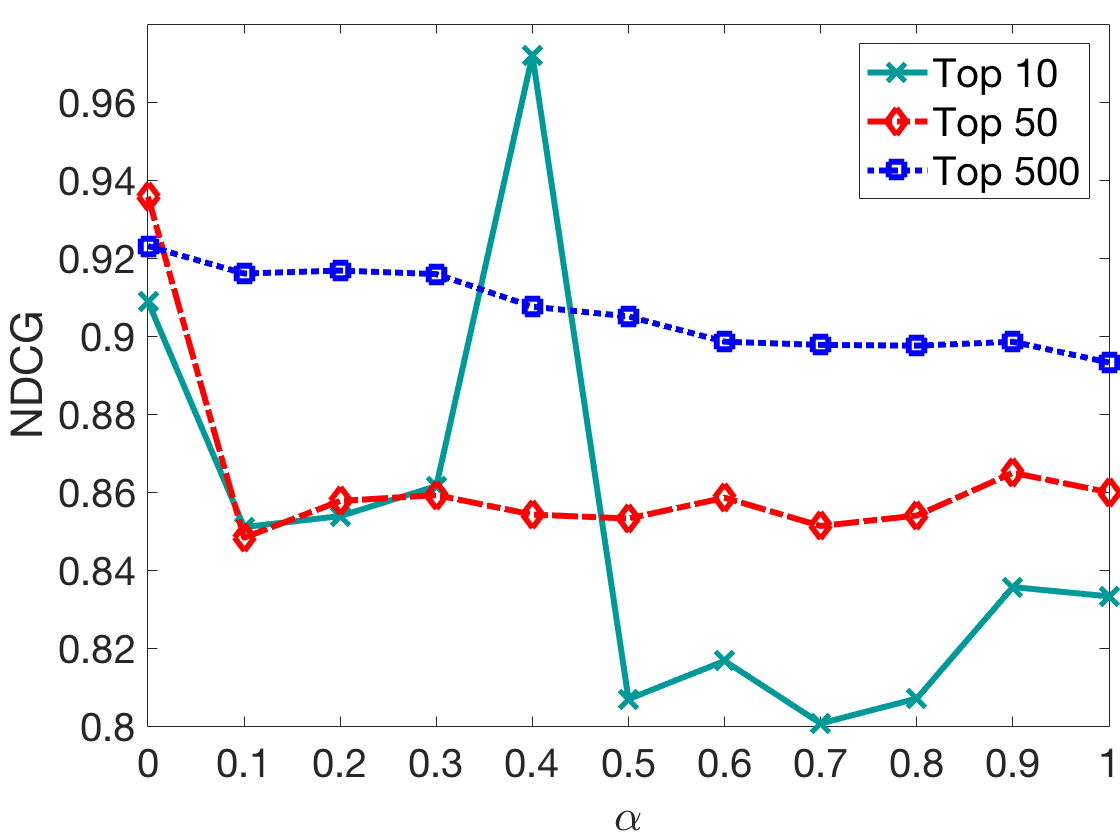}}
	\caption{Parameter analysis of $\alpha$ in linear combination on three datasets. We show Top10, Top50 and Top500 ranking results. $\alpha=0$ means we use motif-based relations while $\alpha=1$ means we use original edge-based relations alone to perform PageRank. Note that for top 10 results with $M_7$ on Epinions, the best performance is achieved with $\alpha=0.9$.}
	\label{fig-vary-alpha}
\end{figure*}

\subsection{Performance of MPR with linear combination}
\label{subsec-exp-res-linear}
In this part, we report the performance of MPR when 3-node motifs and linear combination method are employed. The results are shown in Table~\ref{tb-ndcg-moitf-3-linear}. We show the performance of top10, top50, and top500 results using different algorithms. In addition, we present the results of the baseline methods, anchor motifs, and simple motifs. We highlight in boldface the results of motifs that outperform all baselines, and the best performance of anchor and simple motifs is further underlined. From Table~\ref{tb-ndcg-moitf-3-linear}, we can see that MPR can  outperforms all baselines with different motifs, which demonstrates the effectiveness of adding motif-based relations into PageRank computation for user ranking.  Moreover, the best performance on each dataset is achieved with different types of motifs, meaning that the effectiveness of the motifs is domain specific.

We give a detailed analysis of several observations made from Table~\ref{tb-ndcg-moitf-3-linear} as follows.  
First, when $K=10$, we can see that the NDCGs of BPR on DBLP and Epinions are greater than $0.94$, which is very strong in practice. However, MPR with simple or anchor motifs can still improve the NDCGs further. Specifically, MPR with simple motifs improves the NDCG from $0.9464$ to $0.9920$ on DBLP, and from $0.9777$ to $0.9957$ on Epinions. On Ciao, the performance gain is even greater, which has been improved from $0.8332$ to $0.9905$. The performance gain of anchor motifs are very similar, which is from $0.9464$ to $0.9934$ on DBLP, $0.9777$ to $0.9935$ on Epinions, and from $0.8332$ to $9907$ on Ciao. 

Second, when comparing the performance of all motifs on three datasets, we can observe that on Ciao dataset, there are more motifs outperforming all baselines than those on DBLP and Epinions datasets. Specially, seven simple motifs outperforms all baselines for all results, and for top500 results, all motifs including the 13 motifs improve the NDGCs comparing to all baselines. This might be the fact that the density of social network from Ciao dataset is larger than those of the other two (see Table~\ref{tb-dataset-stat}.), which means the number of motifs is larger. In this sense, it further demonstrate the usefulness of motif for user ranking in social networks.

%Second, when comparing IND, BET, CLO, and BPR, BPR wins only on the Epinions dataset for top10 results. This indicates that the centrality measures are able to characterize the local or global topological structures of a network. WPR is even worse than BPR on DBLP (note that WPR and BPR are the same on Epinions and Ciao). This means that edge weights cannot be chosen arbitrarily. However, by introducing higher-order relations using simple or anchor motifs, MPR can outperform all of the baselines. Therefore, despite the fact that PageRank does not perform well in general compared to the centrality-based methods, the MPR framework can significantly improve the performance of PageRank, which further demonstrates the effectiveness of the motif-based relations.

Finally, when comparing the performance of anchor and simple motifs, we can see that their best performance are very close, except for Ciao, for which the best performance of simple motifs is much better than that of anchor motifs for top50 and top500 ranking lists. The main reason may be that the anchor motifs are ``sub-structure'' of simple motifs as shown in Figure~\ref{fig-motif-example}(b) and \ref{fig-anchor-motif-example}. For example, on DBLP, the best performance of the top10 result is $\hat{M}_{12}$ with anchor motif, and $M_7$ with simple motif. From Figure~\ref{fig-motif-example}(b) and \ref{fig-anchor-motif-example}, we can observe that $\hat{M}_{12}$ and $M_7$ are isomorphic in terms of graph structure. In other words, the higher-order relations encoded in $\hat{M}_{12}$ are already included in $M_7$ (The readers may also refer to the matrix computation methods in Table~\ref{tb-motif-adj-computation} and \ref{tb-anchor-motif-adj-computation}). In fact, the relations encoded in $M_7$ consist of the relations encoded in $\hat{M}_{12}$ and $\hat{M}_{13}$, but the NDCG of $\hat{M}_{13}$ ($0.9879$) is not as good as $\hat{M}_{12}$ ($0.9934$) for the top10 result on DBLP. This may cause the NDCG of $M_7$ ($0.9920$) to be a bit smaller than that of $\hat{M}_{12}$ ($0.9934$).

In summary, these observations show that the linear combination (Eq.~\eqref{eq-linear-combination}) of edge-based and motif-based relations can significantly improve the performance of user ranking comparing to the baselines. Considering that the relations encoded in simple motifs are superset of those in anchor motifs, in the remaining sections, we only show the performance of simple motifs for space limitation.

%\begin{figure*}[h]
%	\subfigure[BPR vs. $M_7$ on DBLP.]{	\includegraphics[width=0.33\textwidth]{top10-trend-case-DBLP}}
%	\subfigure[BPR vs. $M_6$ on Epinions.]{	\includegraphics[width=0.33\textwidth]{top10-trend-case-Epinions}}
%	\subfigure[BPR vs. $M_2$ on Ciao.]{	\includegraphics[width=0.33\textwidth]{top10-trend-case-Ciao}}
%	\caption{The trend of relevance scores of top ten users on all three datasets.}
%	\label{fig-top10-trend-case}
%\end{figure*}

\begin{table*}[]
	\centering
	\caption{NDCGs of the non-linear combination method using seven 3-node simple motifs for top10, top50, top500 users on DBLP, Epinions, and Ciao datasets. The results of motifs which outperform all baselines are in boldface, and the best performance of anchor motifs and simple motifs in each column is further underlined.}
	\label{tb-ndcg-moitf-3-nonlinear}
	\begin{tabular}{c|ccc|ccc|ccc}
		\hline
		& \multicolumn{3}{c|}{DBLP} & \multicolumn{3}{c|}{Epinions} & \multicolumn{3}{c}{Ciao} \\\hline
		TopK & 10 & 50 & 500 & 10 & 50 & 500 & 10 & 50 & 500 \\\hline
		IND & 0.9879 & 0.9639 & 0.9400 & 0.9476 & 0.9563 & 0.9343 & 0.9218 & 0.8651 & 0.9120 \\
		BET & 0.9796 & 0.9710 & 0.9559 & 0.9566 & 0.9559 & 0.9403 & 0.9421 & 0.8961 & 0.8911 \\
		CLO & 0.9875 & 0.9614 & 0.9285 & 0.9308 & 0.9346 & 0.9382 & 0.9021 & 0.9225 & 0.9251 \\
		BPR & 0.9464 & 0.9414 & 0.9527 & 0.9777 & 0.9543 & 0.9365 & 0.8332 & 0.8599 & 0.8932 \\
		WPR & 0.9154 & 0.8871 & 0.9350 & 0.9777 & 0.9543 & 0.9365 & 0.8332 & 0.8599 & 0.8932 \\\hline
		$M_1$ & 0.9859 & \underline{\textbf{0.9716}} & \underline{\textbf{0.9634}} & \textbf{0.9919} & \textbf{0.9629} & \textbf{0.9444} & \underline{\textbf{0.9922}} & \textbf{0.9295} & \underline{\textbf{0.9296}} \\
		$M_2$ & \textbf{0.9890} & 0.9420 & \textbf{0.9560} & \textbf{0.9893} & \textbf{0.9593} & 0.9388 & \textbf{0.9905} & 0.9213 & \textbf{0.9286} \\
		$M_3$ & \textbf{0.9895} & 0.9508 & 0.9528 & \textbf{0.9913} & \underline{\textbf{0.9676}} & 0.9378 & \textbf{0.9551} & 0.8926 & 0.9197 \\
		$M_4$ & \underline{\textbf{0.9979}} & 0.9501 & 0.9521 & \textbf{0.9816} & 0.9543 & 0.9384 & 0.9088 & \underline{\textbf{0.9357}} & 0.9231 \\
		$M_5$ & 0.9653 & 0.9414 & 0.9527 & 0.9777 & 0.9543 & \underline{\textbf{0.9454}} & 0.9055 & 0.9130 & 0.9200 \\
		$M_6$ & 0.9634 & 0.9525 & 0.9588 & \underline{\textbf{0.9960}} & \textbf{0.9621} & 0.9391 & 0.9089 & 0.9058 & 0.9212 \\
		$M_7$ & 0.9869 & 0.9414 & 0.9527 & 0.9777 & \textbf{0.9630} & \textbf{0.9442} & 0.9334 & 0.9079 & 0.9218 \\\hline
	\end{tabular}
\end{table*}

\subsubsection{Significance test of the performance gain}
To figure out the significance of the performance gain in the above section, we run t-test for the results. Specifically, we randomly select 80\% of the nodes from the datasets, and then repeat the experiments for 30 times. For simplicity, we compare BPR, WPR, and MPR, and for MPR, we choose only one motif for each dataset, which is $M_7$ for DBLP and Epinion and $M_4$ for Ciao. We report the average NDCGs of top10, top50, top500 on each dataset. Moreover, we also report the p-values of the baseline methods, i.e., BPR and WPR,  comparing to MPR. The results are shown in Table~\ref{tb-sig-test}.

From Table~\ref{tb-sig-test}, we can see that MPR significantly outperforms BPR and WPR on all three datasets in terms of the NDCGs of top10, top50, top500. This is very consistent with the results in the above section. Moreover, the corresponding p-values show that the performance gain of MPR is statistically significant. It further demonstrates the effectiveness of MPR for the user ranking task in social networks.

\subsubsection{Analysis of $\alpha$}
\label{subsec-alpha-linear}
In Eq.~\eqref{eq-linear-combination}, $\alpha$ is used to balance the edge-based and motif-based relations. $\alpha = 0$ means only motif-based relations are used for authority computation, and $\alpha = 1$ means only edge-based relations are used. In this part, we show how $\alpha$ affects NDCGs on the three datasets. For simplicity, we only show the results of one motif for each dataset, namely, $M_7$ on DBLP and Epinions and $M_4$ on Ciao. The results are shown in Figure~\ref{fig-vary-alpha}.
We can see that the trends are consistent in most cases and the best performance is achieved at some value in $(0,1)$.
It means that the best performance on three datasets is achieved by combining the edge-based and motif-based relations.
It is interesting to note that
the best performance of top500 ranking results on Epinion and Ciao and top50 ranking results on Ciao is achieved at $\alpha = 0$, i.e., using only the motif-based relations.
This again demonstrates that motif-based relations can provide useful information for ranking users in social networks.
Top50 and top500 ranking results are more consistent and the trends of the curves are also similar,
whereas top10 results are more diverse. This may be because the top ten users in the social networks are very prominent and may have different behaviors than the other users.
\vspace{-0.1in}

\begin{figure*}[!ht]
	\subfigure[$M_1$ on DBLP]{	\includegraphics[width=0.33\textwidth]{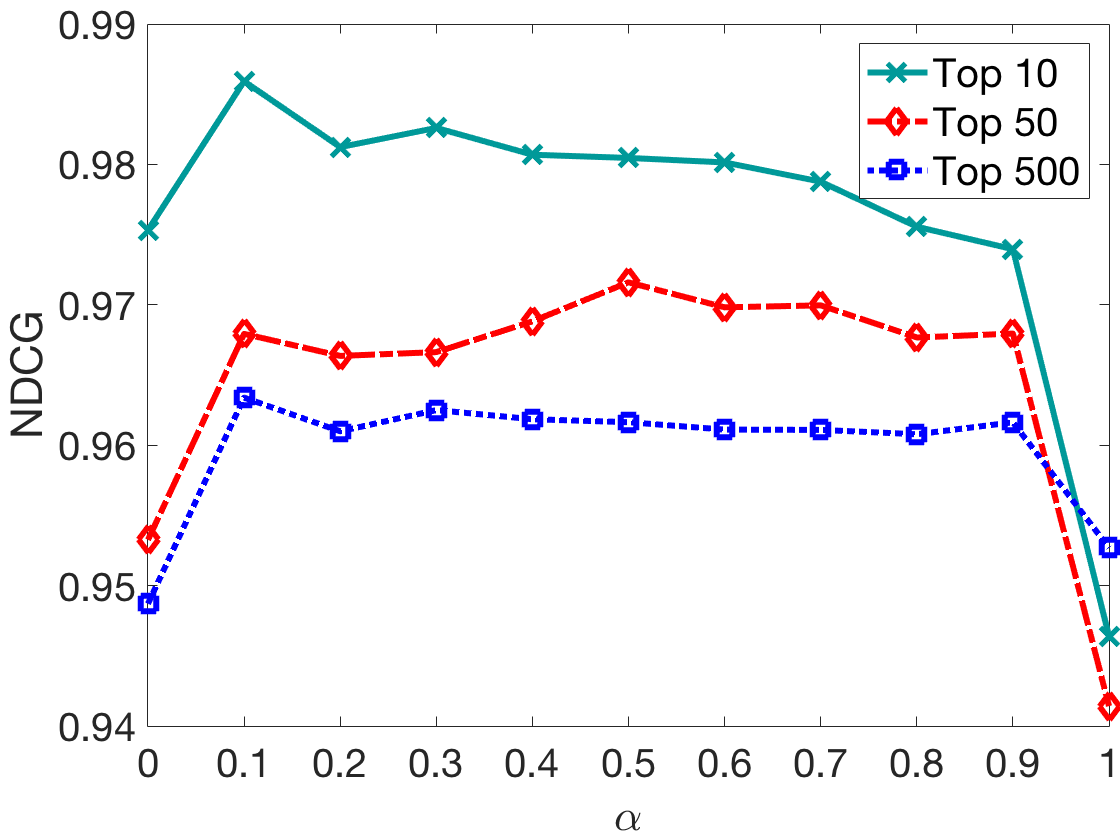}}
	\subfigure[$M_6$ on Epinions]{	\includegraphics[width=0.33\textwidth]{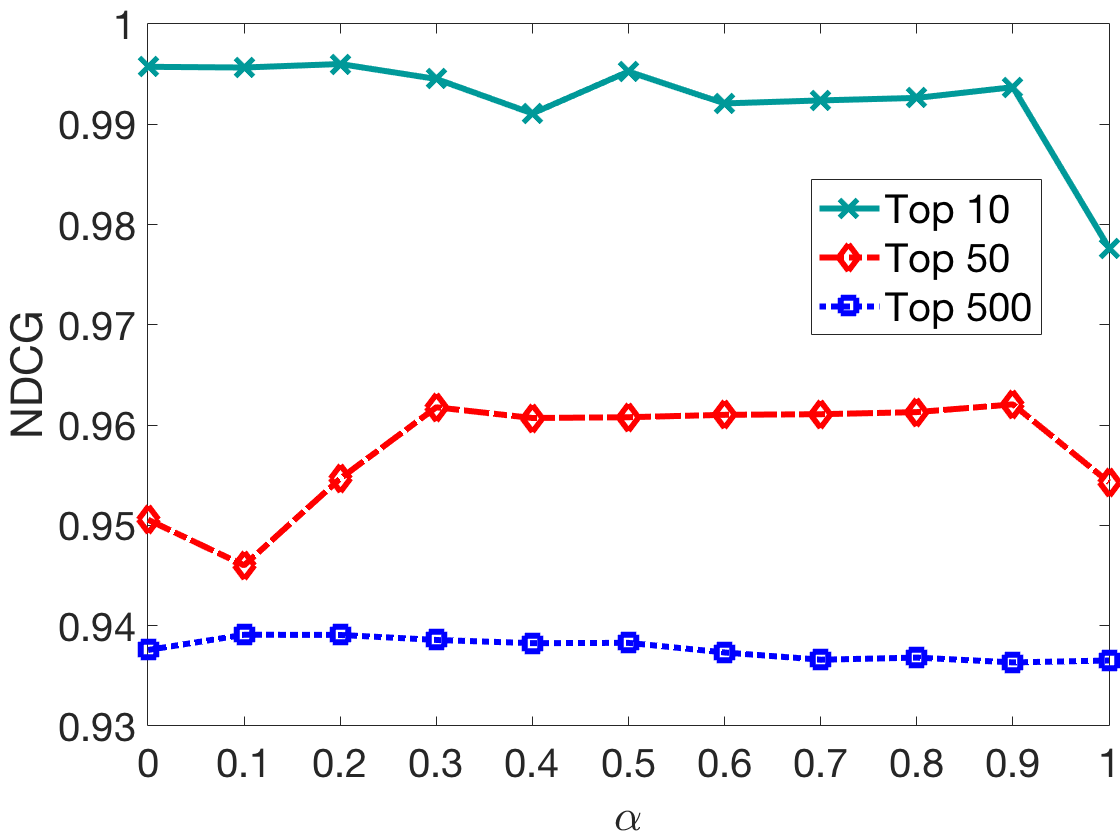}}
	\subfigure[$M_1$ on Ciao]{	\includegraphics[width=0.33\textwidth]{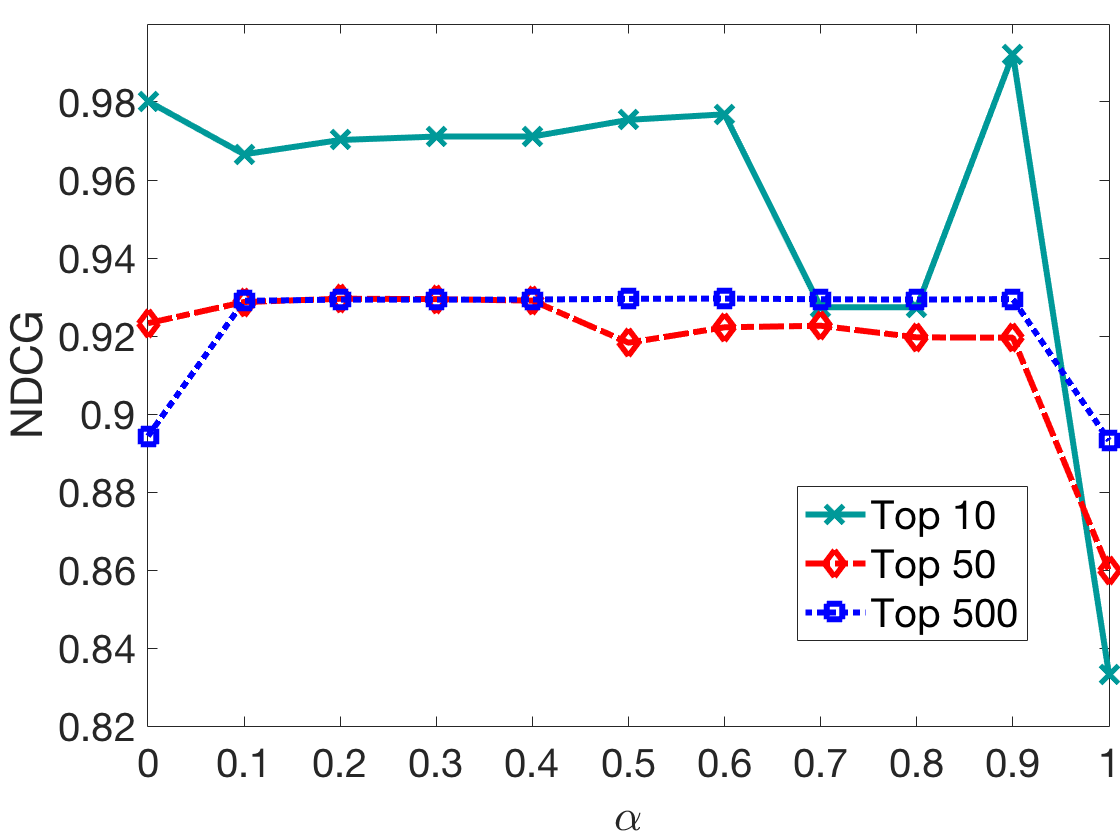}}
	\caption{Parameter analysis of $\alpha$ in non-linear combination on the three datasets. Top10, Top50 and Top500 ranking results are shown. $\alpha=0$ means motif-based relations are used alone to perform PageRank while $\alpha=1$ means the original edge based relations alone are used to perform PageRank.}
	\label{fig-non-linear-vary-alpha}
\end{figure*}

\begin{table*}[]
	\centering
	\caption{NDCGs of the ensemble of the seven 3-node simple motifs. We list the baselines methods for comparison. The best performance of each column is emphasized with bold font. We also list for comparison the best NDCGs of 3-node simple motif, denoted as $M_{best}$.}
	\label{tb-ensemble-motif-3}
	\begin{tabular}{c|ccc|ccc|ccc}
		\hline
		& \multicolumn{3}{c|}{DBLP} & \multicolumn{3}{c|}{Epinions} & \multicolumn{3}{c}{Ciao} \\ \hline
		TopK & 10 & 50 & 500 & 10 & 50 & 500 & 10 & 50 & 500 \\\hline
		IND & 0.9879 & 0.9639 & 0.9400 & 0.9476 & 0.9563 & 0.9343 & 0.9218 & 0.8651 & 0.9120 \\
		BET & 0.9796 & 0.9710 & 0.9559 & 0.9566 & 0.9559 & 0.9403 & 0.9421 & 0.8961 & 0.8911 \\
		CLO & 0.9875 & 0.9614 & 0.9285 & 0.9308 & 0.9346 & 0.9382 & 0.9021 & 0.9225 & 0.9251 \\
		BPR & 0.9464 & 0.9414 & 0.9527 & 0.9777 & 0.9543 & 0.9365 & 0.8332 & 0.8599 & 0.8932 \\
		WPR & 0.9154 & 0.8871 & 0.9350 & 0.9777 & 0.9543 & 0.9365 & 0.8332 & 0.8599 & 0.8932 \\\hline
		$M_{en}$ & \textbf{0.9902} & \textbf{0.9737} & \textbf{0.9641} & \textbf{0.9894}   & \textbf{0.9634} & \textbf{0.9422} & \textbf{0.9741} & \textbf{0.9336} & \textbf{0.9274} \\\hline
		$M_{best}$ & \underline{0.9920} & \underline{0.9766} & \underline{0.9640} & \underline{0.9957}   & \underline{0.9656} & \underline{0.9442} & \underline{0.9905} & \underline{0.9792} & \underline{0.9427} \\\hline
	\end{tabular}
\end{table*}

\subsection{Performance of MPR with non-linear combination}
\label{subsec-exp-res-nonlinear}
In Section~\ref{subsec-exp-res-linear}, we demonstrate the effectiveness of MPR with linear combination (Eq.~\eqref{eq-linear-combination}). In this part, we further show the performance of MPR with non-linear combination method (Eq.~\eqref{eq-nonlinear-combination}). We use 3-node simple motifs and adopt the same experimental settings in Section~\ref{subsec-exp-res-linear}. The results are shown in Table~\ref{tb-ndcg-moitf-3-nonlinear}, and similarly, we highlight in boldface the results better than those of all baselines and further underline the best performance. From Table~\ref{tb-ndcg-moitf-3-nonlinear}, we can see that the non-linear combination method with 3-node simple motifs can also outperform the baselines in NDCG on all three datasets, which also demonstrates the effectiveness of motif for user ranking in social networks. Besides, in Section~\ref{subsec-exp-res-linear}, we observe that the NDCGs can be improved with more types of motifs on Ciao dataset with linear combination method,  however, such phenomenon is not observed with non-linear combination methods. Furthermore, the best performance of linear combination (see Table~\ref{tb-ndcg-moitf-3-linear}) and non-linear combination methods is very close. Therefore, in practice, linear combination is preferred, and we report the results of the linear method in the remaining sections.

Besides the performance comparisons to baseline methods, we also conduct experiments to show the influence of the combination factor, i.e., $\alpha$ in Eq.~\eqref{eq-nonlinear-combination}. According to Eq.~\eqref{eq-nonlinear-combination}, $\alpha$ balances the combination of motif-order and edge-based relations. We vary $\alpha$ in $[0,1]$ in our experiments, with $\alpha=0$ and $\alpha=1$ representing, respectively, motif-order and edge-based relations only. According to Table~\ref{tb-ndcg-moitf-3-nonlinear}, we choose $M_1$ on DBLP and Ciao datasets, and $M_6$ on Epinions dataset. The performance trends of top10, top50, top500 results are shown in Figure~\ref{fig-non-linear-vary-alpha}. We can see that the best performance is achieved when $0 < \alpha < 1$ on all three datasets. It also demonstrates our assumption that the motif-based and edge-based relations are complementary for the user ranking task in social networks.
\vspace{-0.1in}

\subsection{Ensemble of motifs}
\label{subsec-exp-res-ensemble}
In Sections~\ref{subsec-exp-res-linear} and \ref{subsec-exp-res-nonlinear}, we show the performance of each motif separately. It is natural to explore whether the performance can be further improved by an ensemble of all motifs. In this part, we conduct experiments on all three datasets by ensemble of all 3-node simple motifs. Specifically, we construct a new adjacency matrix based on the following:
\begin{equation}
\bW_{M_{en}} = \sum_{k=1}^{7}\frac{1}{7}\bW_{M_k},
\end{equation}
where $\bW_{M_{en}}$ represents the ensemble of the seven 3-node simple motifs. Besides, we choose the linear combination method to combine the edge-based and ensemble of motif-based adjacency matrices, i.e., Eq.~\eqref{eq-linear-combination}, and repeat the experiments in Section~\ref{subsec-exp-res-linear} and \ref{subsec-exp-res-nonlinear}. The results are shown in Table~\ref{tb-ensemble-motif-3}.

From Table~\ref{tb-ensemble-motif-3}, we can see that MPR with $M_{en}$ outperforms baselines methods consistently on all three datasets, demonstrating the effectiveness of ensemble of motifs. When comparing to the best performance of 3-node simple motifs, i.e., $M_1$ to $M_7$, denoted as $M_{best}$, we can see that the NDCGs are very close on DBLP and Epinions, while $M_{best}$ is better than $M_{en}$ in Ciao. It means that MPR cannot get further improvement by the ensemble of motifs and may even impair the performance because the performance of some motifs is not that good, leading to the decrease of NDCGs.

In fact, when looking back at Table~\ref{tb-ndcg-moitf-3-linear}, the simple motifs can also be regarded as an ``ensemble'' of anchor motifs. The performance comparisons are also similar. For example, $M_7$ in Figure~\ref{fig-motif-example}(b) can be regarded as an ensemble of $\hat{M}_{12}$ and $\hat{M}_{13}$ in Figure~\ref{fig-anchor-motif-example}, and on DBLP, the NDCGs of top10 results with $\hat{M}_{12}, \hat{M}_{13}$ and $M_7$ are 0.9934, 0.9879, and 0.9920, respectively. It is $\hat{M}_{13}$ which leads to the inferior NDCG of $M_7$ comparing to $\hat{M}_{12}$. In summary, we can observe that the ensemble of different motifs does not necessarily further improve the performance for the task of user ranking in social networks.

\begin{table*}[]
	\centering
	\caption{NDCGs of 4-node and 5-node motifs for top10, top50, top500 users from DBLP, Epinions, and Ciao datasets. The results of motifs which outperform all baselines are in boldface, and the best performance of anchor motifs and simple motifs in each column is further underlined. The best NDCGs of 3-node motifs from Table~\ref{tb-ndcg-moitf-3-linear} are underlined as comparison.}
	\label{tb-ndcg-motif-4-5}
	\begin{tabular}{c|ccc|ccc|ccc}
		\hline
		& \multicolumn{3}{c|}{DBLP} & \multicolumn{3}{c|}{Epinions} & \multicolumn{3}{c}{Ciao} \\ \hline
		TopK & 10 & 50 & 500 & 10 & 50 & 500 & 10 & 50 & 500 \\ \hline
		IND & 0.9879 & 0.9639 & 0.9400 & 0.9476 & 0.9563 & 0.9343 & 0.9218 & 0.8651 & 0.9120 \\
		BET & 0.9796 & 0.9710 & 0.9559 & 0.9566 & 0.9559 & 0.9403 & 0.9421 & 0.8961 & 0.8911 \\
		CLO & 0.9875 & 0.9614 & 0.9285 & 0.9308 & 0.9346 & 0.9382 & 0.9021 & 0.9225 & 0.9251 \\
		BPR & 0.9464 & 0.9414 & 0.9527 & 0.9777 & 0.9543 & 0.9365 & 0.8332 & 0.8599 & 0.8932 \\
		WPR & 0.9154 & 0.8871 & 0.9350 & 0.9777 & 0.9543 & 0.9365 & 0.8332 & 0.8599 & 0.8932 \\\hline
		$M3_1$ & 0.9616 & 0.9550 & \textbf{0.9615} & \textbf{0.9827}   & \textbf{0.9596} & \textbf{0.9405} & \textbf{0.9751} & \textbf{0.9286} & \textbf{0.9371} \\
		$M3_2$ & \textbf{0.9892} & 0.9509 & \textbf{0.9565} & 0.9507   & \textbf{0.9597} & \textbf{0.9420} & \textbf{0.9839} & \textbf{0.9401} & \textbf{0.9418} \\
		$M3_3$ & \textbf{0.9900} & 0.9580 & \textbf{0.9562} & \textbf{0.9809}   & \textbf{0.9569} & \textbf{0.9407} & \textbf{0.9674} & \textbf{0.9527} & \textbf{0.9399} \\
		$M3_4$ & 0.9716 & 0.9426 & 0.9512 & 0.9747   & \textbf{0.9567} & \textbf{0.9417} & \textbf{0.9726} & \textbf{0.9367} & \underline{\textbf{0.9438}} \\
		$M3_5$ & 0.9762 & 0.9510 & 0.9557 & 0.9780   & \textbf{0.9584} & \textbf{0.9463} & \textbf{0.9627} & \underline{\textbf{0.9634}} & \textbf{0.9327} \\
		$M3_6$ & 0.9569 & 0.9470 & 0.9549 & 0.9515   & \textbf{0.9598} & 0.9386 & \textbf{0.9493} & \textbf{0.9463} & \textbf{0.9368} \\
		$M3_7$ & 0.9712 & \textbf{0.9751} & \underline{\textbf{0.9694}} & \textbf{0.9781}   & \textbf{0.9597} & \textbf{0.9407} & \textbf{0.9444} & \textbf{0.9404} & \textbf{0.9318} \\\hline
		$M4_1$ & 0.9618 & 0.9210 & 0.9451 & 0.9777 & \textbf{0.9575} & \textbf{0.9423} & \textbf{0.9837} & 0.9131 & \textbf{0.9276} \\
		$M4_2$ & \underline{\textbf{0.9909}} & 0.9254 & 0.9425 & 0.9575 & \textbf{0.9567} & \textbf{0.9433} & \underline{\textbf{0.9878}} & 0.9210 & \textbf{0.9297} \\
		$M4_3$ & 0.9498 & 0.9174 & \textbf{0.9570} & \underline{\textbf{0.9846}} & 0.9558 & \underline{\textbf{0.9469}} & \textbf{0.9790} & \textbf{0.9563} & \textbf{0.9305} \\
		$M4_4$ & 0.9686 & \underline{\textbf{0.9769}} & \textbf{0.9605} & \textbf{0.9820} & 0.9558 & \textbf{0.9413} & \textbf{0.9764} & \textbf{0.9277} & \textbf{0.9272} \\
		$M4_5$ & \textbf{0.9907} & 0.9503 & \textbf{0.9568} & \textbf{0.9825} & 0.9559 & 0.9387 & \textbf{0.9764} & \textbf{0.9269} & 0.9271 \\
		$M4_6$ & 0.9533 & 0.9288 & 0.9486 & \textbf{0.9819} & 0.9473 & \textbf{0.9435} & \textbf{0.9764} & \textbf{0.9271} & \textbf{0.9271} \\
		$M4_7$ & 0.9560 & 0.9088 & 0.9427 & 0.9550 & \textbf{0.9584} & \textbf{0.9414} & \textbf{0.9790} & \textbf{0.9456} & \textbf{0.9273} \\\hline
		$M5_1$ & 0.9632 & 0.9128 & 0.9411 & 0.9471 & 0.9562 & \textbf{0.9409} & \textbf{0.9644} & 0.8958 & 0.8968 \\
		$M5_2$ & 0.9629 & 0.9232 & 0.9453 & \textbf{0.9784} & 0.9560 & 0.9390 & \textbf{0.9830} & \textbf{0.9245} & 0.8984 \\
		$M5_3$ & 0.9379 & 0.9209 & 0.9392 & 0.9641 & \underline{\textbf{0.9602}} & 0.9367 & \textbf{0.9696} & 0.9059 & \textbf{0.9271} \\
		$M5_4$ & 0.9742 & \textbf{0.9729} & \textbf{0.9567} & \textbf{0.9823} & 0.9559 & 0.9384 & 0.9248 & 0.9052 & 0.8979 \\
		$M5_5$ & \textbf{0.9895} & 0.9227 & 0.9421 & \textbf{0.9820} & 0.9558 & 0.9355 & 0.9347 & 0.8809 & 0.8958 \\
		$M5_6$ & 0.9454 & 0.9007 & 0.9386 & \textbf{0.9822} & 0.9539 & \textbf{0.9436} & 0.8665 & \textbf{0.9294} & 0.9006 \\
		$M5_7$ & 0.9587 & 0.8996 & 0.9386 & \textbf{0.9787} & 0.9552 & 0.9387 & \textbf{0.9570} & 0.8827 & 0.8943\\\hline
		M3 & \underline{0.9920} & \underline{0.9766} & \underline{0.9640} & \underline{0.9957} & \underline{0.9656} & \underline{0.9442} & \underline{0.9905} & \underline{0.9792} & \underline{0.9427}\\\hline
	\end{tabular}
\end{table*}

\subsection{The performance of 4-node and 5-node motifs}
\label{subsec-exp-res-4-5-motif}
In this part, we show the performance of several 4-node and 5-node motifs. Since there are 199 isomorphic 4-node and 9364 5-node subgraphs~\cite{kashtan2004efficient}, it is impossible to enumerate all of the motifs and report their performance in our experiments. Moreover, not all subgraphs are motifs according to the definition of motif. Therefore, instead of enumerating all possible motifs in our experiments, we only report the performance of top seven motifs based on Z-score and concentration, obtained from mFinder\footnote{https://www.weizmann.ac.il/mcb/UriAlon/download/network-motif-software}, which is implemented according to the sampling method~\cite{kashtan2004efficient}.

\begin{figure*}
	\centering
	\includegraphics[width=0.9\textwidth]{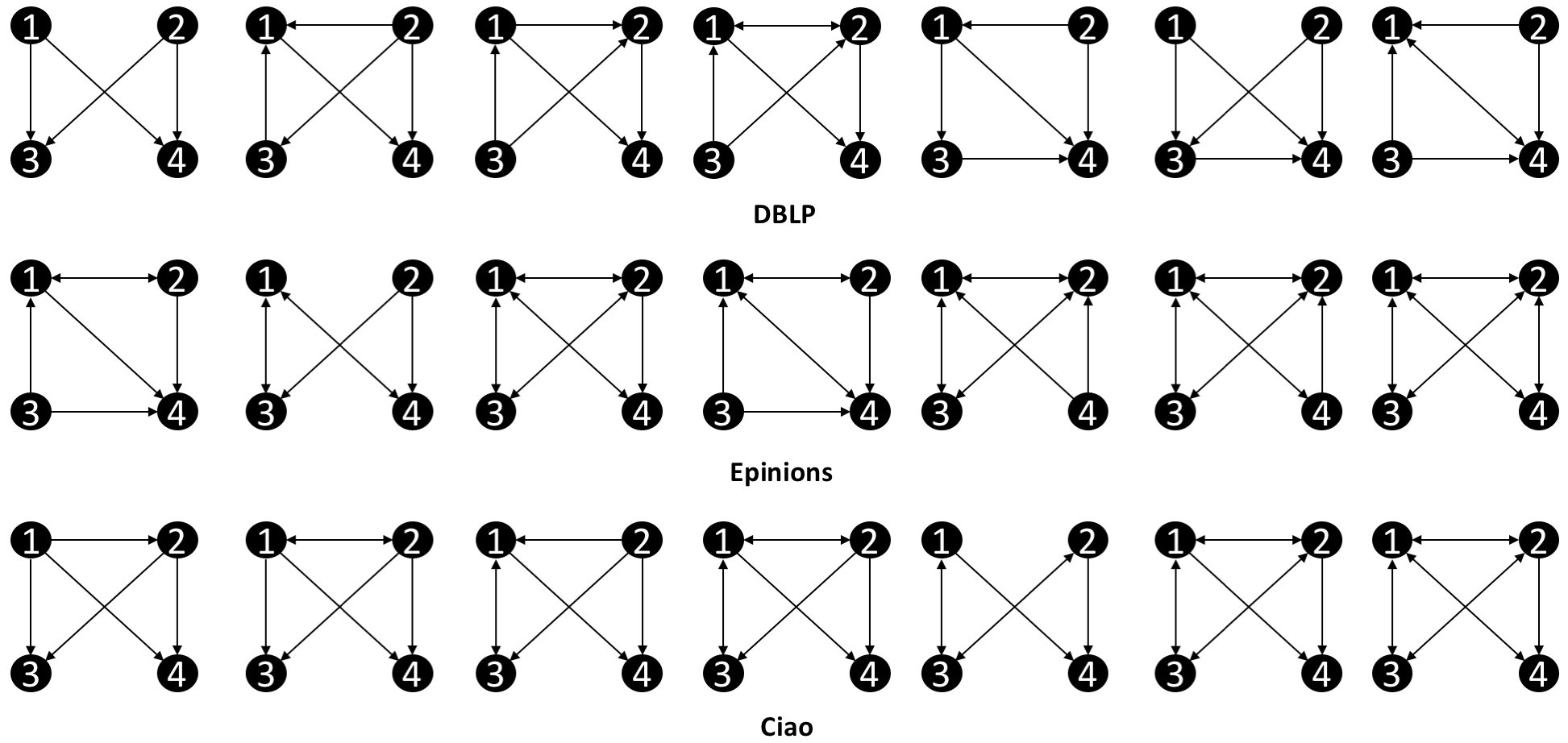}
	\centering
	\caption{The seven 4-node motifs used on the experiments, which are denoted as $M4_1,..., M4_7$ from left to right. Note that these motifs are detected by the method in~\cite{kashtan2004efficient}.}
	\label{fig-4node-motif}  
\end{figure*}

\begin{figure*}
	\centering
	\includegraphics[width=0.9\textwidth]{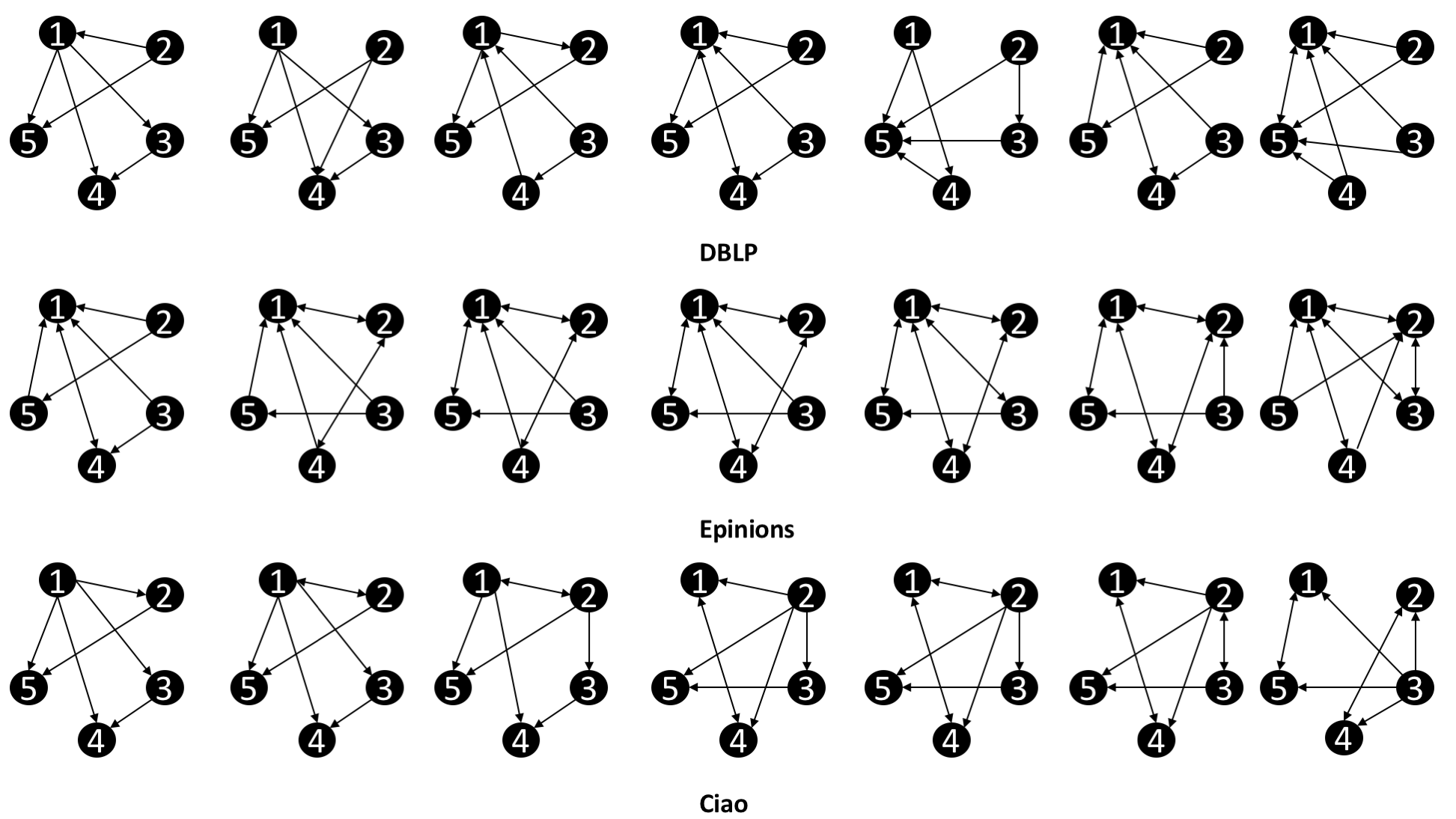}
	\centering
	\caption{The seven 5-node motifs used in the experiments, which are denoted as $M5_1,..., M5_7$ from left to right. Note that these motifs are detected by the method in~\cite{kashtan2004efficient}.}
	\label{fig-5node-motif}  
\end{figure*}

Specifically, on all three datasets, we first run mFinder to sample subgraphs and find the top seven motifs. The number of sampled subgraphs is set to 1,000,000.\footnote{Note that in the mFinder documentation, the authors suggested the number to be larger than 50,000 for a good estimation, and the larger the better. Here, we set it to 1,000,000 according to the computational power of our machines.} The seven 4-node and 5-node motifs of each dataset are shown in Figure~\ref{fig-4node-motif} and \ref{fig-5node-motif}, respectively. Besides, mFinder also returns all instances of each motif. We then can construct the motif-based adjacency matrix from the instances according to Algorithm~\ref{alg-motif-4-5-adj-computation}. Finally, we repeat the same experiments as those for 3-node motifs on all three datasets. Note that we choose the linear combination method. Here we also report the performance of 3-node motifs by Algorithm~\ref{alg-motif-4-5-adj-computation} as comparison.

%\subsubsection{Performance analysis}

The NDCGs of different motifs are shown in Table~\ref{tb-ndcg-motif-4-5}. Three interesting observations can be obtained from the table. First, having more nodes in a motif does not lead to better performance. In most cases, the best performance is obtained with 4-node motifs. This may be attributed to the noise in motifs with a large number of nodes that degrades the performance. Second, the performance gain of large motifs is smaller than that of 3-node motifs, which aligns with the first observation and research finding in social network that triangles are indicators for strong relations in social networks~\cite{simmel1908sociology,granovetter1977strength}. Third, same as that in Section~\ref{subsec-alpha-linear}, the NDCGs can be improved with more types of motifs on Ciao dataset. Taking all into consideration, the experimental results further demonstrate the effectiveness of the proposed MPR framework.
\vspace{-0.1in}
%\subsection{Analysis of $\alpha$}
%In this part, we show the impact of $\alpha$ in the supervised learning task. Similar to the settings in Section~\ref{subsec-alpha-unsupervise}, we show the performance trend of the motif obtaining better results. The results are shown in Figure~\ref{fig-supervised-vary-alpha}.
%
%\begin{figure*}
%	       \centering
%	       \subfigure[PLCC@DBLP.]{ \includegraphics[width=0.32\textwidth]{DBLP_PLCC}}
%	       \subfigure[NDCG@DBLP.]{ \includegraphics[width=0.32\textwidth]{DBLP_NDCG}}
%	       \subfigure[RMSE@DBLP.]{ \includegraphics[width=0.32\textwidth]{DBLP_RMSE}}\\
%	       \subfigure[PLCC@Epinions.]{     \includegraphics[width=0.32\textwidth]{Epinions_PLCC}}
%	       \subfigure[NDCG@Epinions.]{     \includegraphics[width=0.32\textwidth]{Epinions_NDCG}}
%	       \subfigure[RMSE@Epinions.]{     \includegraphics[width=0.32\textwidth]{Epinions_RMSE}}\\
%	       \subfigure[PLCC@Ciao.]{ \includegraphics[width=0.32\textwidth]{Ciao_PLCC}}
%	       \subfigure[NDCG@Ciao.]{ \includegraphics[width=0.32\textwidth]{Ciao_NDCG}}
%	       \subfigure[RMSE@Ciao.]{ \includegraphics[width=0.32\textwidth]{Ciao_RMSE}}
%	       \centering
%	       \caption{Performance trending by varying $\alpha$.}
%	       \label{fig-supervised-vary-alpha}
%\end{figure*}

\subsection{Discussions}
Based on the experimental results, we demonstrate the effectiveness of the proposed MPR framework on ranking users in social networks. In this part, we discuss several questions in the following.
\begin{itemize}
	\item In Section~\ref{subsec-exp-res-linear}, \ref{subsec-exp-res-nonlinear}, and \ref{subsec-exp-res-4-5-motif}, we show the performance of each motif, separately. Clearly the performance of each type of motif varies, and it is difficult to decide the best motif in different fields. In this paper, we demonstrate the importance of considering local structures when ranking users in social networks with PageRank, and design a data-driven pipeline to incorporate the motif-based higher-order relations. When applying to new domains, the MPR framework and pipeline can be efficiently adopted then.
	\item When comparing linear and non-linear combination methods, in our experimental results, as described in Section~\ref{subsec-alpha-linear}, \ref{subsec-exp-res-nonlinear} and \ref{subsec-exp-res-4-5-motif}, the linear combination method is preferred due to the larger positive gains. However, considering the proposed combination methods are simple (Eq.~\eqref{eq-linear-combination} and Eq.~\ref{eq-nonlinear-combination}), it is worthwhile to determine more complex methods to combine the first-order and higher-order relations, which can be a future direction. 
	\item In Section~\ref{subsec-exp-res-ensemble}, we further present the results of ensemble of seven 3-node motifs, observing that it is not necessarily to improve the performance comparing to a single motif since not all motifs can lead to positive gains. It is worthwhile to explore more ensembles of different types of motifs, while naive search is definitely infeasible in terms of time cost. Therefore, we leave it for future work to explore more efficient methods for ensemble of different motifs.
\end{itemize}
\vspace{-0.2in}

\section{Learning task with motif}
\label{sec-learning-exp}
In this section, to further show the usefulness of motif-based higher-order relations, we design a learning task, authority prediction, in social networks. Generally speaking, we extract features for each node from different adjacency matrices, and train a learning model to predict the authority of a node. In this way, we can compare the prediction abilities of features from different adjacency matrices, including the original one $\bW$ and the combined ones $\bH_{M_k}$ according to~\eqref{eq-linear-combination}, thus evaluating the effectiveness of motif-based relations for authority prediction. The datasets used are DBLP, Epinions and Ciao in Section~\ref{subsec-unsup-exp-settings}, and the authority values, i.e., H-index on DBLP, and the trustworthiness score on Epinions and Ciao, are used as labels. For simplicity, we choose seven 3-node simple motifs in Figure~\ref{fig-motif-example}(b) for this experiments, thus the combined adjacency matrices are denoted as $\bH_{M_1},\cdots,\bH_{M_7}$.

Specifically, the experiment consists of the following steps:
\begin{itemize}
	\item Extract, respectively, four features from original and combined adjacency matrices, which are: in-degree, out-degree, betweenness, and closeness.
	\item Random sample 1500 and 500 nodes from the graph as train data and test data, respectively.
	\item Train a Support Vector Regression (SVR)~\cite{cortes1995support} model using the authority values as labels.
	\item After training, we use the trained SVR to predict the authority of nodes in test data, and then compare them with ground-truth.
\end{itemize}
For the combined adjacency matrices, $\bH_{M_1},\cdots,\bH_{M_7}$, we obtain different ones according to Eq.~\eqref{eq-linear-combination} by varying $\alpha$ as $\{0.1,0.2,\cdots,1.0\}$, and report the best performance. To evaluate the accuracy of the predicted authority, we use two metrics: NDCG and RMSE. For NDCG, we rank the nodes in test data with the predicted authority and compute the NDCG according to Eq.~\eqref{eq-ndcg}, i.e., $NDCG_{500}$. For RMSE, it is calculated as $\sqrt{\frac{1}{|\bar{\Omega}|} \sum_{i \in \bar{\Omega}}{(\by_i - \hat{\by}_i)^2}}$, where $\bar{\Omega}$ is the test data. $\by_i$ and $\hat{\by}_i$ are the ground-truth and predicted authority values for node $u_i$, respectively. For these two metrics, larger NDCG means better performance while smaller RMSE means better performance. We repeat the process for 10 times, with the test data fixed and different train data. Finally we report the mean of the results from 10 rounds, which are shown in Table~\ref{tb-supervise-res}. Again, we highlight in boldface all results better than baseline and further underline the best one.

From Table~\ref{tb-supervise-res}, we can see that the performance in terms of, NDCG and RMSE with $\bH_{M_1},\cdots,\bH_{M_7}$ are significantly and consistently better than those with $\bW$. It means the features extracted from the combined adjacency matrices are more effectiveness than those from the original adjacency matrix. Despite there are two RMSEs ($\bH_{M_5}$ and $\bH_{M_7}$) on Ciao dataset larger than that of $\bW$, the NDCGs with these two motifs are still higher than that of $\bW$. Moreover, the performance gains are all from $\alpha \in (0, 1)$. 
Therefore, these results clearly demonstrate the usefulness of the motif-based higher-order relations for authority prediction in social networks, and that the edge-based and motif-based relations are complementary for this task as well.
\begin{table}[]
	\centering
	\caption{Experimental results of authority prediction. The evaluation metrics include  NDCG and RMSE. The results of motifs which outperform all baselines are in boldface, and the best performance of anchor motifs and simple motifs in each column is further underlined.}
	\label{tb-supervise-res}
\begin{tabular}{c|cc|cc|cc}
	\toprule
 & \multicolumn{2}{c|}{DBLP} & \multicolumn{2}{c|}{Epinions} & \multicolumn{2}{c}{Ciao} \\\hline
Metrics & NDCG & RMSE & NDCG & RMSE & NDCG & RMSE \\\midrule
$W$ & 0.8351 & 13.3930 & 0.9194 & 0.3821 & 0.9158 & 0.4430 \\
$\bH_{M_1}$ & \textbf{0.8746} & \textbf{12.4593} & {\ul \textbf{0.9420}} & \textbf{0.3331} & \textbf{0.9394} & {\ul \textbf{0.3457}} \\
$\bH_{M_2}$ & \textbf{0.8737} & \textbf{12.7931} & \textbf{0.9365} & \textbf{0.3507} & {\ul \textbf{0.9411}} & \textbf{0.3810} \\
$\bH_{M_3}$ & \textbf{0.8689} & \textbf{12.7512} & \textbf{0.9386} & \textbf{0.3478} & \textbf{0.9402} & \textbf{0.3726} \\
$\bH_{M_4}$ & {\ul \textbf{0.8931}} & \textbf{12.0888} & \textbf{0.9384} & {\ul \textbf{0.3149}} & \textbf{0.9395} & \textbf{0.3590} \\
$\bH_{M_5}$ & \textbf{0.8684} & \textbf{11.8914} & \textbf{0.9377} & \textbf{0.3467} & \textbf{0.9293} & 0.4654 \\
$\bH_{M_6}$ & \textbf{0.8732} & {\ul \textbf{11.3917}} & \textbf{0.9378} & \textbf{0.3522} & \textbf{0.9367} & \textbf{0.4262} \\
$\bH_{M_7}$ & \textbf{0.8709} & \textbf{12.2407} & \textbf{0.9368} & \textbf{0.3587} & \textbf{0.9301} & 0.4442\\\bottomrule
\end{tabular}
\end{table}

\section{Conclusion and Future Work}
\label{sec-conclusion}
In this paper, we propose the motif-based PageRank (MPR) for user ranking in social networks, which incorporates higher-order relations. To combine the motif-based and edge-based relations in PageRank computatiopn, we propose a linear and a non-linear combination method, respectively. We conduct extensive experiments on three real-world datasets with different types of motifs. All the experimental results demonstrate that MPR can significantly improve the performance of ranking users in social networks.

For future work, we point out here three possible directions. First, it remains to be seen whether more complex combination methods can be designed. Second, it will be interesting to explore ensembles of more motifs, beyond the 3-node motifs in this paper. Finally, it is worthwhile to explore whether motif can enhance other graph-based tasks.

\section{Acknowledgment}
This paper was supported by HKUST-WeChat WHAT Lab, China 973 Fundamental R\&D Program (No. 2014CB340304), and the Research Grants Council HKSAR GRF (No.16215019 and No. 26206717). We also thank the anonymous reviewers for their valuable comments and suggestions that help improve the quality of this manuscript.

%\clearpage
\bibliographystyle{IEEEtran}
\bibliography{draft}

\begin{thebibliography}{10}
\providecommand{\url}[1]{#1}
\csname url@samestyle\endcsname
\providecommand{\newblock}{\relax}
\providecommand{\bibinfo}[2]{#2}
\providecommand{\BIBentrySTDinterwordspacing}{\spaceskip=0pt\relax}
\providecommand{\BIBentryALTinterwordstretchfactor}{4}
\providecommand{\BIBentryALTinterwordspacing}{\spaceskip=\fontdimen2\font plus
\BIBentryALTinterwordstretchfactor\fontdimen3\font minus
  \fontdimen4\font\relax}
\providecommand{\BIBforeignlanguage}[2]{{%
\expandafter\ifx\csname l@#1\endcsname\relax
\typeout{** WARNING: IEEEtran.bst: No hyphenation pattern has been}%
\typeout{** loaded for the language `#1'. Using the pattern for}%
\typeout{** the default language instead.}%
\else
\language=\csname l@#1\endcsname
\fi
#2}}
\providecommand{\BIBdecl}{\relax}
\BIBdecl

\bibitem{benson2016higher}
A.~R. Benson, D.~F. Gleich, and J.~Leskovec, ``Higher-order organization of
  complex networks,'' \emph{Science}, vol. 353, no. 6295, pp. 163--166, 2016.

\bibitem{SongCHT07}
X.~Song, Y.~Chi, K.~Hino, and B.~L. Tseng, ``Identifying opinion leaders in the
  blogosphere,'' in \emph{CIKM}, 2007, pp. 971--974.

\bibitem{TangSWY09}
J.~Tang, J.~Sun, C.~Wang, and Z.~Yang, ``Social influence analysis in
  large-scale networks,'' in \emph{KDD}, 2009, pp. 807--816.

\bibitem{xiang2013pagerank}
B.~Xiang, Q.~Liu, E.~Chen, H.~Xiong, Y.~Zheng, and Y.~Yang, ``Pagerank with
  priors: {A}n influence propagation perspective,'' in \emph{IJCAI}, 2013, pp.
  2740--2746.

\bibitem{WangWTZC15}
Y.~Wang, X.~Wang, J.~Tang, W.~Zuo, and G.~Cai, ``Modeling status theory in
  trust prediction,'' in \emph{AAAI}, 2015, pp. 1875--1881.

\bibitem{page1999pagerank}
L.~Page, S.~Brin, R.~Motwani, and T.~Winograd, ``The pagerank citation ranking:
  {B}ringing order to the web,'' Stanford InfoLab, Tech. Rep., 1999.

\bibitem{xing2004weighted}
W.~Xing and A.~Ghorbani, ``Weighted pagerank algorithm,'' in \emph{Proceedings.
  Second Annual Conference on Communication Networks and Services Research,
  2004.}\hskip 1em plus 0.5em minus 0.4em\relax IEEE, 2004, pp. 305--314.

\bibitem{radicchi2009diffusion}
F.~Radicchi, S.~Fortunato, B.~Markines, and A.~Vespignani, ``Diffusion of
  scientific credits and the ranking of scientists,'' \emph{Physical Review E},
  vol.~80, no.~5, p. 056103, 2009.

\bibitem{zyczkowski2010citation}
K.~{\.Z}yczkowski, ``Citation graph, weighted impact factors and performance
  indices,'' \emph{Scientometrics}, vol.~85, no.~1, pp. 301--315, 2010.

\bibitem{ding2011applying}
Y.~Ding, ``Applying weighted pagerank to author citation networks,''
  \emph{Journal of the American Society for Information Science and Technology
  (JASIST)}, vol.~62, no.~2, pp. 236--245, 2011.

\bibitem{milo2002network}
R.~Milo, S.~Shen-Orr, S.~Itzkovitz, N.~Kashtan, D.~Chklovskii, and U.~Alon,
  ``Network motifs: {S}imple building blocks of complex networks,''
  \emph{Science}, vol. 298, no. 5594, pp. 824--827, 2002.

\bibitem{simmel1908sociology}
G.~Simmel, ``Sociology: investigations on the forms of sociation,''
  \emph{Duncker \& Humblot, Berlin Germany}, 1908.

\bibitem{granovetter1977strength}
M.~S. Granovetter, ``The strength of weak ties,'' in \emph{Social
  networks}.\hskip 1em plus 0.5em minus 0.4em\relax Elsevier, 1977, pp.
  347--367.

\bibitem{zhao2018ranking}
H.~Zhao, X.~Xu, Y.~Song, D.~L. Lee, Z.~Chen, and H.~Gao, ``Ranking users in
  social networks with higher-order structures,'' in \emph{AAAI}, 2018.

\bibitem{gleich2015pagerank}
D.~F. Gleich, ``Pagerank beyond the web,'' \emph{SIAM Review}, vol.~57, no.~3,
  pp. 321--363, 2015.

\bibitem{ding2011topic}
Y.~Ding, ``Topic-based pagerank on author cocitation networks,'' \emph{Journal
  of the American Society for Information Science and Technology (JASIST)},
  vol.~62, no.~3, pp. 449--466, 2011.

\bibitem{liben2007link}
D.~Liben-Nowell and J.~Kleinberg, ``The link-prediction problem for social
  networks,'' \emph{Journal of the American Society for Information Science and
  Technology (JASIST)}, vol.~58, no.~7, pp. 1019--1031, 2007.

\bibitem{ugander2013subgraph}
J.~Ugander, L.~Backstrom, and J.~Kleinberg, ``Subgraph frequencies: Mapping the
  empirical and extremal geography of large graph collections,'' in \emph{WWW},
  2013, pp. 1307--1318.

\bibitem{granovetter1973strength}
M.~S. Granovetter, ``The strength of weak ties,'' \emph{American journal of
  sociology (AJS)}, vol.~78, no.~6, pp. 1360--1380, 1973.

\bibitem{rotabi2017detecting}
R.~Rotabi, K.~Kamath, J.~Kleinberg, and A.~Sharma, ``Detecting strong ties
  using network motifs,'' in \emph{WWW Companion}, 2017, pp. 983--992.

\bibitem{wang2014identification}
P.~Wang, J.~L{\"u}, and X.~Yu, ``Identification of important nodes in directed
  biological networks: A network motif approach,'' \emph{PloS one}, vol.~9,
  no.~8, p. e106132, 2014.

\bibitem{prvzulj2007biological}
N.~Pr{\v{z}}ulj, ``Biological network comparison using graphlet degree
  distribution,'' \emph{Bioinformatics}, vol.~23, no.~2, pp. e177--e183, 2007.

\bibitem{sporns2004motifs}
O.~Sporns and R.~K{\"o}tter, ``Motifs in brain networks,'' \emph{PLoS biology},
  vol.~2, no.~11, p. e369, 2004.

\bibitem{paranjape2017motifs}
A.~Paranjape, A.~R. Benson, and J.~Leskovec, ``Motifs in temporal networks,''
  in \emph{WSDM}, 2017, pp. 601--610.

\bibitem{ahmed2015efficient}
N.~K. Ahmed, J.~Neville, R.~A. Rossi, and N.~Duffield, ``Efficient graphlet
  counting for large networks,'' in \emph{ICDM}, 2015, pp. 1--10.

\bibitem{jha2015path}
M.~Jha, C.~Seshadhri, and A.~Pinar, ``Path sampling: A fast and provable method
  for estimating 4-vertex subgraph counts,'' in \emph{WWW}, 2015, pp. 495--505.

\bibitem{wang2016minfer}
P.~Wang, J.~C. Lui, D.~Towsley, and J.~Zhao, ``Minfer: A method of inferring
  motif statistics from sampled edges,'' in \emph{ICDE}, 2016, pp. 1050--1061.

\bibitem{han2016waddling}
G.~Han and H.~Sethu, ``Waddling random walk: Fast and accurate mining of motif
  statistics in large graphs,'' in \emph{ICDM}, 2016, pp. 181--190.

\bibitem{stefani2017triest}
L.~D. Stefani, A.~Epasto, M.~Riondato, and E.~Upfal, ``Tri{\`e}st: Counting
  local and global triangles in fully dynamic streams with fixed memory size,''
  \emph{ACM Transactions on Knowledge Discovery from Data (TKDD)}, vol.~11,
  no.~4, p.~43, 2017.

\bibitem{pinar2017escape}
A.~Pinar, C.~Seshadhri, and V.~Vishal, ``Escape: {E}fficiently counting all
  5-vertex subgraphs,'' in \emph{WWW}, 2017, pp. 1431--1440.

\bibitem{yin2017local}
H.~Yin, A.~R. Benson, J.~Leskovec, and D.~F. Gleich, ``Local higher-order graph
  clustering,'' in \emph{KDD}, 2017, pp. 555--564.

\bibitem{zhang2017structinf}
J.~Zhang, J.~Tang, Y.~Zhong, Y.~Mo, J.~Li, G.~Song, W.~Hall, and J.~Sun,
  ``Struct{I}nf: {M}ining structural influence from social streams.'' in
  \emph{AAAI}, 2017, pp. 73--80.

\bibitem{bianchini2005inside}
M.~Bianchini, M.~Gori, and F.~Scarselli, ``Inside pagerank,'' \emph{ACM
  Transactions on Internet Technology (TOIT)}, vol.~5, no.~1, pp. 92--128,
  2005.

\bibitem{bensonsupplementary}
A.~R. Benson, D.~F. Gleich, and J.~Leskovec, ``Supplementary materials for
  higher-order organization of complex networks,'' \emph{Science}, 2016.

\bibitem{kashtan2004efficient}
N.~Kashtan, S.~Itzkovitz, R.~Milo, and U.~Alon, ``Efficient sampling algorithm
  for estimating subgraph concentrations and detecting network motifs,''
  \emph{Bioinformatics}, vol.~20, no.~11, pp. 1746--1758, 2004.

\bibitem{tang2008arnetminer}
J.~Tang, J.~Zhang, L.~Yao, J.~Li, L.~Zhang, and Z.~Su, ``Arnet{M}iner:
  {E}xtraction and mining of academic social networks,'' in \emph{KDD}, 2008,
  pp. 990--998.

\bibitem{tang-etal12a}
J.~Tang, H.~Gao, and H.~Liu, ``m{T}rust: {D}iscerning multi-faceted trust in a
  connected world,'' in \emph{WSDM}, 2012, pp. 93--102.

\bibitem{tang-etal12b}
J.~Tang, H.~Gao, H.~Liu, and A.~Das~Sarma, ``e{T}rust: {U}nderstanding trust
  evolution in an online world,'' in \emph{KDD}, 2012, pp. 253--261.

\bibitem{jarvelin2002cumulated}
K.~J{\"a}rvelin and J.~Kek{\"a}l{\"a}inen, ``Cumulated gain-based evaluation of
  ir techniques,'' \emph{ACM Transactions on Information Systems (TOIS)},
  vol.~20, no.~4, pp. 422--446, 2002.

\bibitem{hirsch2005index}
J.~E. Hirsch, ``An index to quantify an individual's scientific research
  output,'' \emph{Proceedings of the National academy of Sciences (PNAS)}, vol.
  102, no.~46, pp. 16\,569--16\,572, 2005.

\bibitem{freeman1977set}
L.~C. Freeman, ``A set of measures of centrality based on betweenness,''
  \emph{Sociometry}, pp. 35--41, 1977.

\bibitem{sabidussi1966centrality}
G.~Sabidussi, ``The centrality index of a graph,'' \emph{Psychometrika},
  vol.~31, no.~4, pp. 581--603, 1966.

\bibitem{cortes1995support}
C.~Cortes and V.~Vapnik, ``Support-vector networks,'' \emph{Machine learning},
  vol.~20, no.~3, pp. 273--297, 1995.

\end{thebibliography}

\clearpage
\begin{IEEEbiography}[{\includegraphics[width = 1\textwidth]{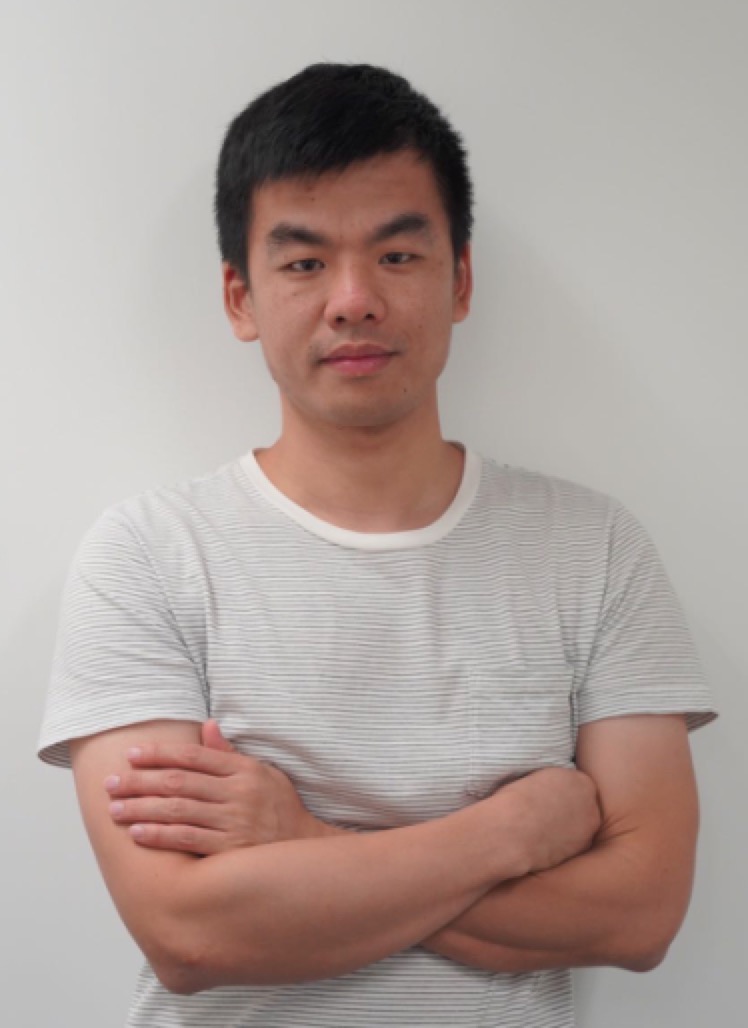}}]{Huan Zhao} is now a research scientist in 4Paradigm Inc., China. He obtained his Ph.D. degree from the the Department of Computer Science and Engineering in the Hong Kong University of Science and Technology in Jan. 2019. His research interests focus on recommender systems and data mining. He received the BSc degree from Beijing University of Posts and Telecommunications (BUPT) in 2012. He is a member of IEEE.
\end{IEEEbiography}

%\vspace{-0.2in}

\begin{IEEEbiography}[{\includegraphics[width = 1\textwidth]{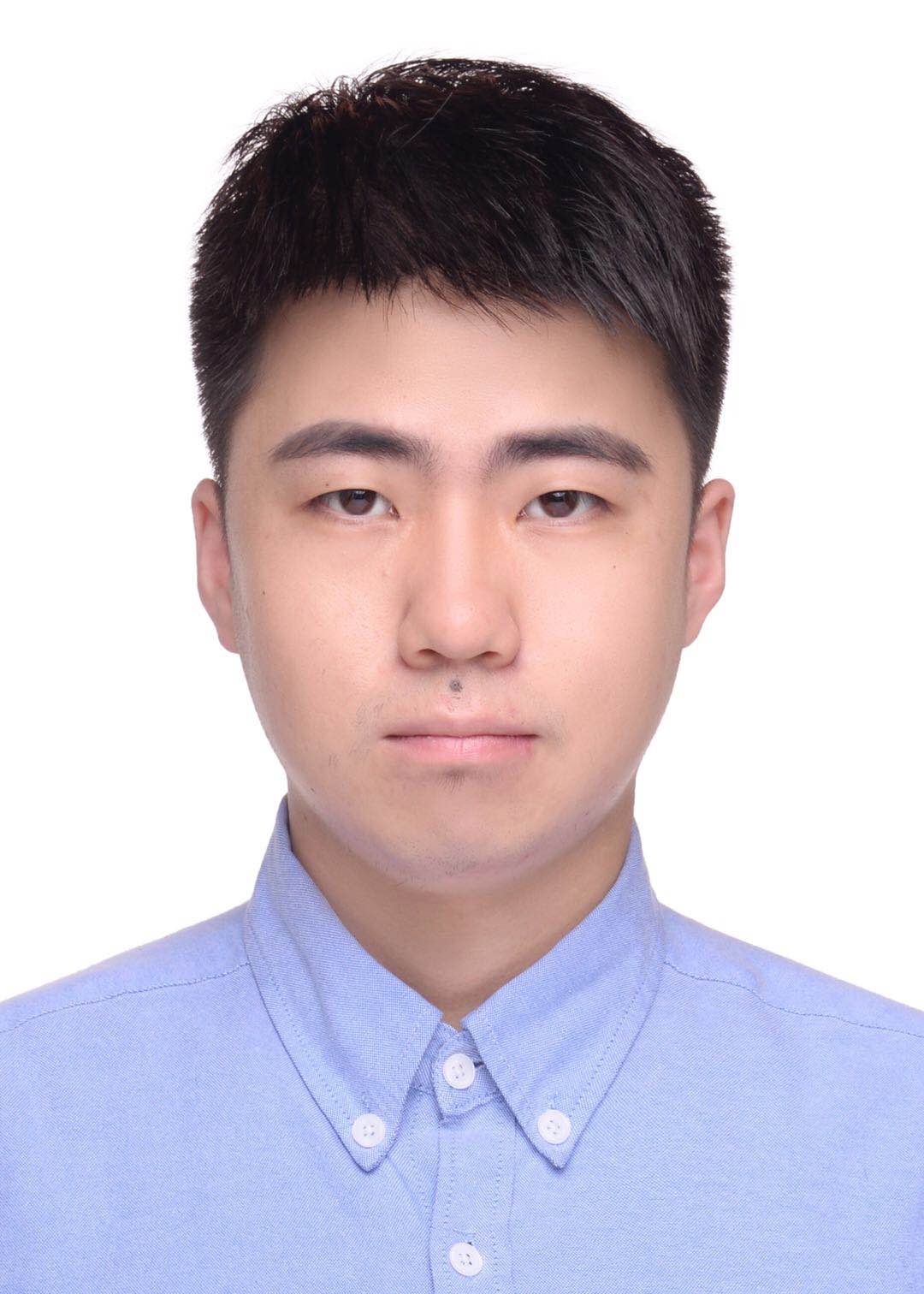}}]{Xiaogang Xu} received the B.Sc. degree from Zhejiang University, China, in 2018. His research interests include computer vision, image processing and machine learning.
\end{IEEEbiography}
%\vspace{-0.2in}
\begin{IEEEbiography}[{\includegraphics[width = 1\textwidth]{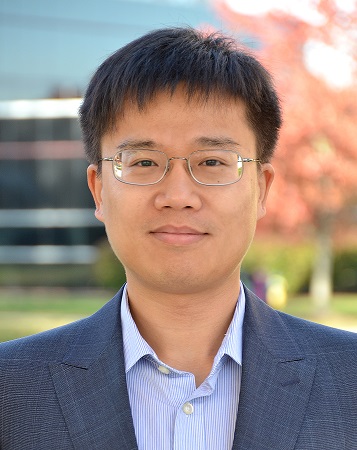}}]{Yangqiu Song} is now an assistant professor at Department of CSE with a joint appointment at Math Department at HKUST. He was an assistant professor at Lane Department of CSEE at WVU  during 2015 and 2016, a post-doc researcher at UIUC during 2013 and 2015, a post-doc researcher at HKUST and visiting researcher at Huawei Noah's Ark Lab, Hong Kong during 2012 and 2013, an associate researcher at Microsoft Research Asia during 2010 and 2012, and a staff researcher at IBM Research-China  during 2009 and 2010. He received my B.E. and Ph.D. degree from Tsinghua University, China, in July 2003 and January 2009. He also worked as an intern at Google during 2007 and 2008 and at IBM Research-China during 2006 and 2007. His research interests are knowledge graph, text mining, information extraction and machine learning.
\end{IEEEbiography}

%\vspace{-0.2in}

\begin{IEEEbiography}[{\includegraphics[width = 1\textwidth]{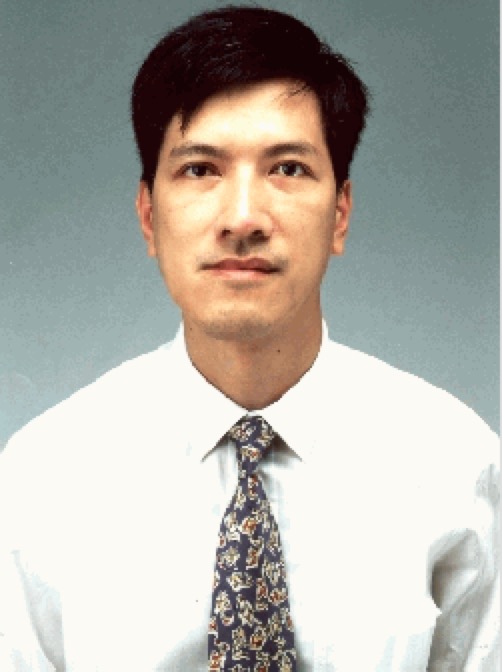}}]{Dik Lun Lee} received the BSc degree in electronics from the Chinese University of Hong Kong, and the MS and PhD degrees in computer science from the University of Toronto, Canada. 
He is currently a professor in the Department of Computer Science and Engineering at the Hong Kong University of Science and Technology. He was an associate professor in the Department of Computer Science and Engineering at the Ohio State University. His research interests include information retrieval, search engines, mobile computing, and pervasive computing.
\end{IEEEbiography}

\begin{IEEEbiography}[{\includegraphics[width = 1\textwidth]{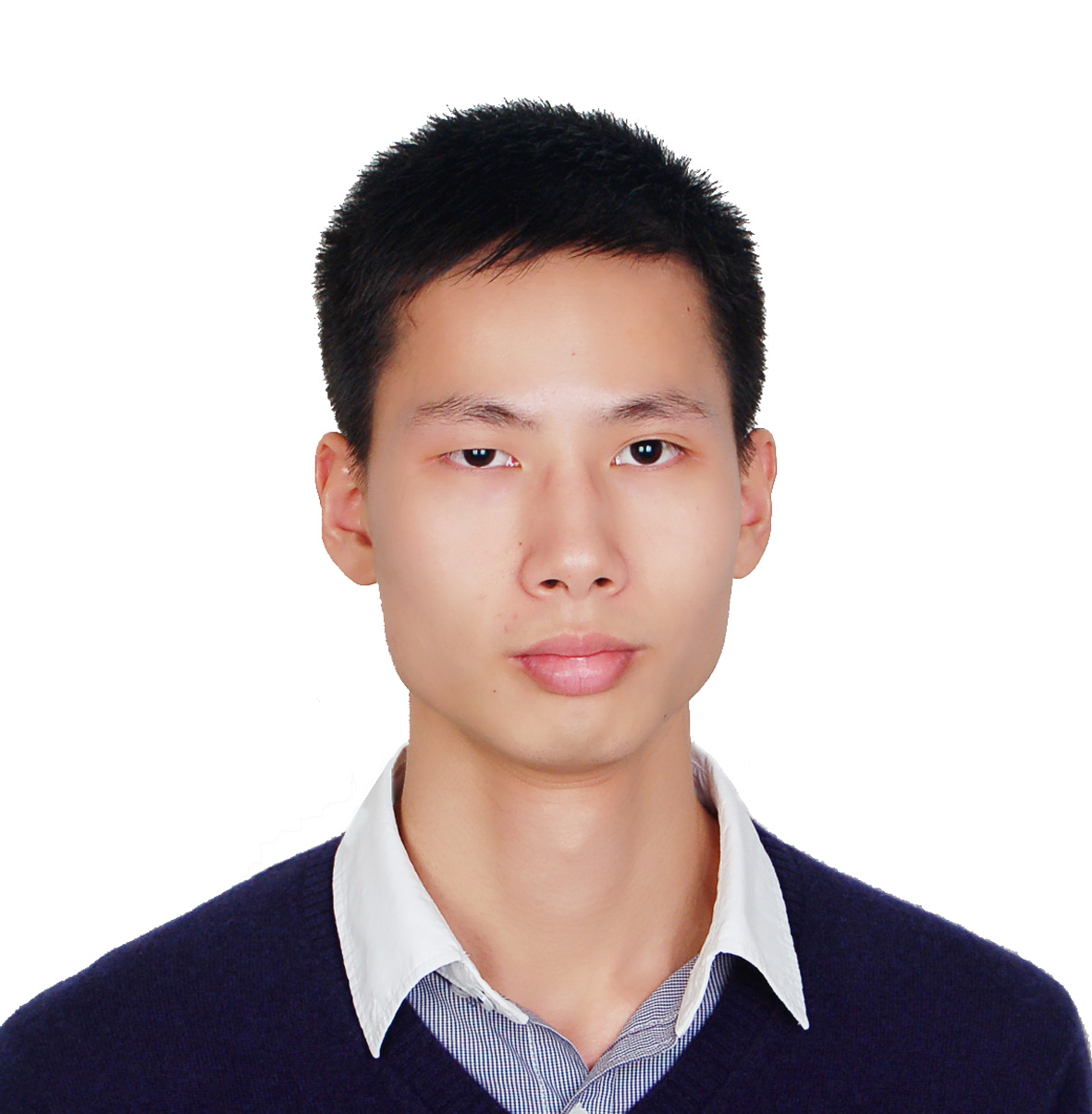}}]{Zhao Chen} received the MSc degree from Harbin Institute of Technology Shenzhen Graduate School in 2016. 
Currently, he is working for Tencent and his research interests focus on social media data mining.
\end{IEEEbiography}
%\vspace{-7in}
\begin{IEEEbiography}[{\includegraphics[width = 1\textwidth]{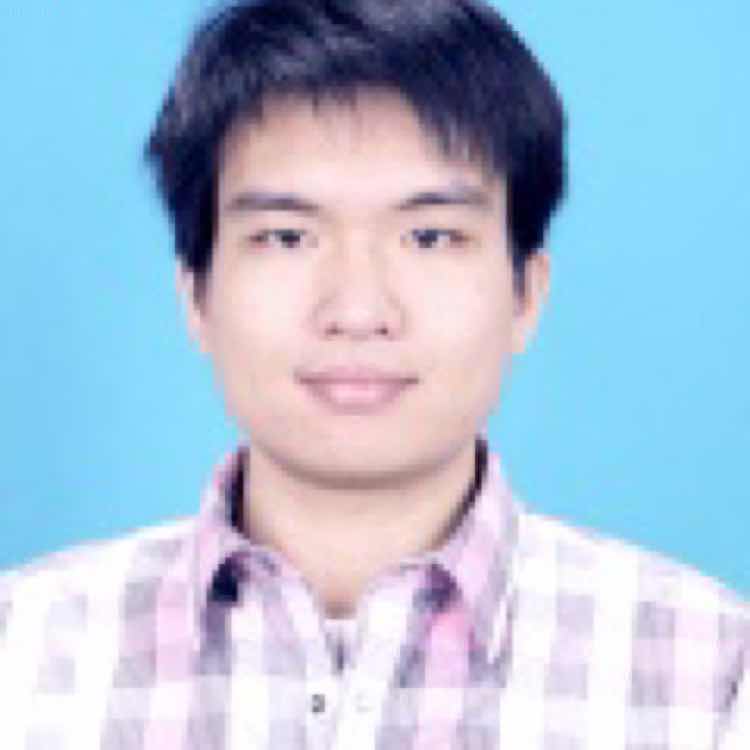}}]{Han Gao} received the MSc and BSc degrees from School of Mathematics, Sun Yat-Sen University. Currently, he is working as a researcher in Tencent Technology (Shenzhen) Co., Ltd. His research interests focus on social network and data mining.
\end{IEEEbiography}

\end{document}